\documentclass[prd,twocolumn,nofootinbib,superscriptaddress]{revtex4-2}
\usepackage{bm}
\usepackage{amsmath}
\usepackage{amssymb}
\usepackage{graphicx}
\usepackage{subfigure}
\usepackage{hyperref}
\usepackage{color}
\usepackage{comment}
\usepackage{ulem}
\usepackage{CJK}
\hypersetup{
	colorlinks=true,
	linkcolor=red,
	citecolor=blue,
}

\allowdisplaybreaks[2]

\usepackage{graphicx}
\usepackage{dcolumn}
\usepackage{bm}
\begin{document}

\preprint{APS/123-QED}

\title{The influence of the mean anomaly on the dynamical quantities of binary black hole mergers in eccentric orbits}

\author{Hao Wang}
\email{wanghao_zju@zju.edu.cn}
\affiliation{Institute for Astronomy, School of Physics, Zhejiang University, 310058 Hangzhou, China}
\affiliation{Department of Astronomy, School of Physics, Huazhong University of Science and Technology, Wuhan 430074, China}

\author{Bin Liu}
\email{liubin23@zju.edu.cn}
\affiliation{Institute for Astronomy, School of Physics, Zhejiang University, 310058 Hangzhou, China}

\author{Yuan-Chuan Zou}
\email{zouyc@hust.edu.cn}
\affiliation{Department of Astronomy, School of Physics, Huazhong University of Science and Technology, Wuhan 430074, China}

\author{Qing-Wen Wu}
\email{qwwu@hust.edu.cn}
\affiliation{Department of Astronomy, School of Physics, Huazhong University of Science and Technology, Wuhan 430074, China}

\date{\today}

\begin{abstract}
In studies of binary black hole (BBH) mergers in eccentric orbits, the mean anomaly, traditionally regarded as less significant than eccentricity, has been thought to encode only the orbital phase, leading to the assumption that it exerts minimal influence on the dynamics of eccentric mergers.
In a previous investigation, we identified consistent oscillations in dynamical quantities peak luminosity $L_{\text{peak}}$, remnant mass $M_{\text{rem}}$, spin $\alpha_{\text{rem}}$, and recoil velocity $V_{\text{rem}}$—in relation to the initial eccentricity $e_0$. These oscillations are associated with integer orbital cycles within a phenomenological framework. In this paper, we aim to explore the underlying physical nature of these oscillations through gravitational waveforms.
Our examination of remnant mass and spin reveals that while the initial ADM mass $M_{\mathrm{ADM}}$ and orbital angular momentum $L_0$ exhibit gradual variations with $e_0$, the radiated energy $E_{\text{rad}}$ and angular momentum $L_{\text{rad}}$ display oscillatory patterns akin to those observed in $M_{\text{rem}}$ and $\alpha_{\text{rem}}$. By decomposing the waveforms into three distinct phases—inspiral, late inspiral to merger, and ringdown, we demonstrate that these oscillations persist across all phases, suggesting a common origin.
Through a comparative analysis of $E_{\text{rad}}$ and $L_{\text{rad}}$ derived from numerical relativity (NR), post-Newtonian (PN) waveforms, and orbital-averaged PN fluxes during the inspiral phase, we identify the initial mean anomaly $l_0$ as the source of the observed oscillations. This effect, which is averaged out in orbital-averaged flux calculations, significantly influences $M_{\text{rem}}$, $\alpha_{\text{rem}}$ and $V_{\text{rem}}$, with its impact increasing as $e_0$ rises.
Further, we find that by continuously varying $l_0$ within the parameter space $[0,2\pi]$, we can construct an envelope that encompasses the original oscillations of these radiative quantities, indicating that the oscillations originate from the specific initial condition $l_0$.
We synthesize and analyze the relationships between dynamical quantities and mass ratio for orbital BBH mergers, integrating data from both eccentric and circular orbits. Our findings emphasize that eccentricity and mean anomaly induce oscillations and ranges in dynamical quantities relative to circular orbits. We interpolate the maximum and minimum values of the dynamical quantities to delineate the vicinities of these quantities for eccentric orbits compared to circular orbits. Notably, the vicinities intensify with higher mass ratios ($q=m_1/m_2\leq1$, $m_1$ and $m_2$ are component masses of the BBH) for $M_{\text{rem}}$, $\alpha_{\text{rem}}$ and $L_{\text{peak}}$, reaching maximum effects for $q \approx 1/3$ on $V_{\text{rem}}$.
By quantifying the residual deviations relative to circular orbits, we highlight significant differences between the vicinities and the polynomial modeling of circular orbits, underscoring the importance of this effect, which cannot be overlooked and has a broad impact.

\end{abstract}
\maketitle

\section{Introduction}
The epoch of regular gravitational wave detection began in 2015, signifying a momentous leap forward in astrophysical observations \cite{LIGOScientific:2016aoc}. Currently, the ground-based gravitational wave detectors, including LIGO \cite{LIGOScientific:2014qfs}, Virgo \cite{VIRGO:2014yos}, and KAGRA \cite{KAGRA:2018plz}, have collectively detected nearly one hundred binary compact object mergers \cite{LIGOcollaboration}.

The primary focus of contemporary gravitational wave detections pertains to circular orbit BBH mergers, primarily attributed to the circularization effect induced by gravitational radiation \cite{Peters:1963ux,Peters:1964zz}. Nonetheless, within star-dense environments like globular clusters \cite{Miller:2002pg,Gultekin:2005fd,OLeary:2005vqo,Rodriguez:2015oxa,Samsing:2017xmd,Rodriguez:2017pec,Rodriguez:2018pss,Park:2017zgj} and galactic nuclei \cite{Gondan:2020svr,Antonini:2012ad,Kocsis:2011jy,Hoang:2017fvh,Gondan:2017wzd,Samsing:2020tda,Tagawa:2020jnc}, dynamical interactions involving two-body \cite{East:2012xq}, three-body \cite{Naoz:2012bx,VanLandingham:2016ccd,Silsbee:2016djf,Blaes:2002cs,Antognini:2013lpa,Stephan:2016kwj,Katz:2011hn,Seto:2013wwa,Liu:2018nrf,Liu:2019tqr,Liu:2024oyc}, four-body \cite{Zevin:2018kzq,Liu:2018vzk} systems, and beyond, can instigate the formation of BBH systems with eccentric orbits. The origin of circular orbit BBH mergers, whether arising from isolated binary evolution or dynamical interactions, remains ambiguous. However, the identification of an eccentric orbit BBH merger undeniably indicates the formation of a dynamically evolved system.
The event GW190521 \cite{LIGOScientific:2020iuh} stands as a notable example, being identified as a probable outcome of an eccentric orbit BBH merger \cite{Gayathri:2020coq,Romero-Shaw:2020thy}, indicative of dynamical formation processes at play.

Numerous techniques exist for resolving the dynamics of BBH mergers, including Effective One Body (EOB) \cite{Buonanno:1998gg,Damour:2001tu}, post-Newtonian \cite{Blanchet:2013haa}, black hole perturbation theory \cite{Teukolsky:1973ha}, and numerical relativity \cite{Pretorius:2005gq}. Among these, NR stands out as the most precise method capable of fully elucidating the strong field dynamics of BBH systems. From NR simulations, we derive essential information like gravitational waveforms and dynamical quantities such as peak luminosity, mass, spin, recoil velocity, etc., which serve as direct inputs for constructing gravitational wave templates and modeling the properties of remnant black hole in astrophysical contexts. 

Modeling dynamical quantities involves establishing connections between the initial parameters and final parameters of BBH systems through various methodologies. Over the last decade, NR has made significant strides, progressing from initial equal-mass nonspinning BBH configurations to encompass spin-aligned, spin-precessing, and unequal-mass BBH systems, and evolving from circular orbit BBH mergers to include eccentric orbits \cite{Duez:2018jaf,Campanelli:2005dd,Baker:2005vv}, and has reached a very high precision. Several NR collaborations, such as Simulating eXtreme Spacetimes (SXS) \cite{Mroue:2013xna,Boyle:2019kee}, Rochester Institute of Technology (RIT) \cite{Healy:2017psd,Healy:2019jyf,Healy:2020vre,Healy:2022wdn}, bi-functional adaptive mesh (BAM) \cite{Hamilton:2023qkv,Bruegmann:2006ulg,Husa:2007hp}, and MAYA \cite{Jani:2016wkt,Ferguson:2023vta}, have conducted extensive simulations across these diverse scenarios. These findings establish favorable conditions for modeling dynamical quantities.
During this period, numerous studies have endeavored to model the relationships between the initial parameters of NR simulated BBH systems and crucial dynamical quantities like peak luminosity $L_{\text{peak}}$, mass $M_{\text{rem}}$, spin $\alpha_{\text{rem}}$, and recoil velocity $V_{\text{rem}}$, employing a range of different approaches. However, the bulk of these modeling efforts have predominantly focused on circular orbit BBH mergers. For instance, some Refs. \cite{Hofmann:2016yih,Lousto:2009mf,Buonanno:2007sv,Lousto:2009ka,Rezzolla:2007rz} have correlated the remnant mass and spin with the energy and angular momentum of the innermost stable circular orbit and mass ratio, while others have used PN \cite{Blanchet:2005rj}, EOB \cite{Damour:2006tr}, and closed limit approximation \cite{Sopuerta:2006wj} methods to model the remnant recoil velocity. Additionally, certain studies have employed polynomial fitting of initial spin and mass ratio to model the remnant mass, spin, recoil velocity, and peak luminosity \cite{Taylor:2020bmj,Healy:2016lce,Healy:2014yta,Lousto:2013wta,Zlochower:2015wga,Jimenez-Forteza:2016oae,Rezzolla:2007xa,Ferguson:2019slp,Koppitz:2007ev,Campanelli:2007cga,Gonzalez:2006md,Hemberger:2013hsa,Carullo:2023kvj}, while others \cite{Varma:2018aht} have utilized non-analytical methods like Gaussian process regression to model the remnant mass, spin, and recoil velocity. The references cited above focus on modeling the dynamical quantities of BBH mergers in circular orbits. Several of these approaches rely on analytical approximations and neglect the strong-field effects during the merger phase, thereby introducing significant limitations.

There has been a notable absence of modeling efforts focused on dynamical quantities in BBH mergers with eccentric orbits. This scarcity primarily stems from the lack of extensive NR simulations for BBH mergers in eccentric orbits. However, in recent years, collaborations like SXS, MAYA, and RIT have expanded their BBH simulations to include eccentric orbits \cite{Mroue:2013xna,Boyle:2019kee,Healy:2017psd,Healy:2019jyf,Healy:2020vre,Healy:2022wdn,Jani:2016wkt,Ferguson:2023vta}.

In terms of modeling dynamical quantities in eccentric BBH mergers, only a single Ref. \cite{Islam:2021mha} has attempted to utilize Gaussian process regression to model the mass and spin of remnants based on twenty sets of eccentric orbit waveforms from SXS. However, this modeling effort is hindered by a limited sample size, and due to the relatively low eccentricities in these NR simulations, the properties of the remnants closely resemble those from circular orbits, lacking universal significance.

Notably, the fourth data release \cite{Healy:2022wdn} from RIT offers a wealth of data with sufficiently dense (close to continuous change) initial eccentricities and a broad coverage of parameter spaces in NR simulations. This dataset \cite{RITBBH} presents a valuable opportunity for in-depth exploration of the characteristics of dynamical quantities in BBH mergers with eccentric orbits.

To date, only a limited number of studies have examined dynamical quantities in eccentric orbits, with most being qualitative in nature. Sopuerta et al. \cite{Sopuerta:2006et} provided a PN perspective, suggesting that in the low eccentricity regime, the recoil velocity $V_\text{rem}$ scales as $\propto\left(1+e_0\right)$, where $e_0$ is the initial eccentricity defined in the reference. This offers analytical insight into the impact of low eccentricity scenarios. Nonetheless, this approach has not been cross-validated against NR simulations and remains confined to the PN regime. 
Radia et al. \cite{Radia:2021hjs} and Sperhake et al. \cite{Sperhake:2019wwo} demonstrated that non-zero eccentricity can lead to a significant increase in the recoil velocity $V_\text{rem}$, potentially by up to 25\% compared to quasi-circular orbits. This indicates that eccentricity indeed influences $V_\text{rem}$ at certain phases. However, due to limited data, the underlying mechanism behind the increase of 
$V_\text{rem}$ in the presence of eccentricity remains poorly understood.
Hinder et al. \cite{Hinder:2007qu} conducted a series of numerical simulations focusing on high eccentricity scenarios. They investigated the dependence of the remnant spin and mass on eccentricity and concluded that the orbit effectively circularizes once the eccentricity drops below 0.4. Higher eccentricities were observed to significantly affect dynamical properties. Here as well, no explanation was provided for the influence of eccentricity on the remnant mass and spin.
In a related study, Sperhake et al. \cite{Sperhake:2007gu} investigated the evolution from inspiral to plunge in eccentric BBH mergers, covering a broad range of eccentricities from 0 to 1. They examined the correlation between eccentricity and emitted energy and angular momentum, notably observing that as the system approached the critical transition from orbiting to plunging, the remnant spin parameter $\alpha_\text{rem}$ peaked at 0.724. This suggests that the effects of eccentricity reach their maximum at a specific critical point. Although this study revealed the existence of such a critical effect and identified a general trend of dynamical quantities with eccentricity, the limited NR data prevented the discovery of the oscillatory structure that we report below.
Radia et al. \cite{Radia:2021hjs} performed numerous nonspinning eccentric numerical simulations with mass ratios of $q=1/2$, $1/3$, and $2/3$. Their research focused on the remnant recoil velocity, revealing intriguing oscillatory behavior. Additionally, their figures displayed noticeable oscillations in the remnant's spin and radiated energy, although these quantities were not the primary focus of their study. That work reported oscillations in $V_\text{rem}$ with eccentricity and attributed them to changes in the infall direction during the BBH merger, but the explanation remained purely phenomenological.
Ref. \cite{Nee:2025zdy} examined the influence of the mean anomaly on dynamical quantities within the EOB framework, but the study remains incomplete as it does not incorporate full NR data for eccentric BBH mergers.

Earlier investigations have progressively unveiled the impact of eccentricity on dynamical quantities; however, the lack of NR data has hindered the comprehensiveness of their research.
In our previous work \cite{Wang:2023vka}, we conducted a comprehensive analysis of the relationship between the dynamical quantities peak luminosity $L_{\text{peak}}$, remnant mass $M_{\text{rem}}$, spin $\alpha_{\text{rem}}$, and recoil velocity $V_{\text{rem}}$ and initial eccentricity $e_0$ for BBH mergers on eccentric orbits, utilizing all available data from RIT. The initial eccentricity $e_0$ is derived from RIT's empirical formula $e_0=2 \epsilon-\epsilon^2$ at the initial moment of the simulation, where $\epsilon$ ranges from 0 to 1 \cite{Healy:2022wdn}. Our findings revealed a gradual oscillation in these dynamical quantities as the initial eccentricity increases, ultimately reaching extrema at specific eccentricity values.
This suggested a deeper structural relationship between eccentricity and dynamical quantities. We attempted to explain this phenomenon phenomenologically by associating the integer orbital cycle number of gravitational waves with the peaks and valleys of the oscillations in dynamical quantities. However, this approach does not elucidate the underlying physical nature of the oscillations induced by eccentricity.

In this paper, we aim to elucidate the nature of oscillations in dynamical quantities from the perspective of waveforms. Building upon our previous work comparing eccentric PN and NR waveforms in Ref. \cite{Wang:2024jro}, we focus on waveforms that exhibit oscillatory patterns during the inspiral phase. By calculating the radiated energy and angular momentum and comparing these results with PN predictions, we identify that the oscillations arise from the influence of the initial mean anomaly $l_0$. Here, $l_0$ denotes the mean anomaly at the initial time of the waveform, obtained by fitting the PN waveform to the NR waveform, and is defined by $l=n\left(t-t_0\right) = l(t)-l_0$, where $n$ is the mean motion and $t_0$ is the initial time \cite{Wang:2024jro}.
This parameter emerges as a crucial factor in BBH mergers, alongside initial eccentricity, and has often been overlooked in prior analyses.

This paper is structured as follows. In Sec. \ref{sec:II}, we introduce the NR data of BBH mergers, which are detailed in Sec. \ref{sec:II:A}, along with a comparison between PN and NR waveforms discussed in Sec. \ref{sec:II:B}.
In Sec. \ref{sec:III}, we present the envelope and vicinities resulting from the effect of the mean anomaly. Finally, in Sec. \ref{sec:IV}, we provide a comprehensive summary and outlook.
Throughout this work, we adopt geometric units where $G=c=1$. The reduced mass is calculated as $\mu = m_1 m_2 / M$, while the symmetric mass ratio is defined as $\eta = \mu / M$.

\section{Method}\label{sec:II}
\subsection{NR data}\label{sec:II:A}
We review the origin and composition of the NR simulation data concerning BBH mergers utilized in our study. The data are sourced from the SXS \cite{SXSBBH} and RIT \cite{RITBBH} NR catalogs.

The SXS collaboration employs a multi-domain spectral method \cite{Szilagyi:2009qz,Kidder:1999fv}, incorporating a first-order iteration of the generalized harmonic formulation \cite{Hemberger:2012jz,Pretorius:2004jg,Garfinkle:2001ni} of Einstein's equations with constraint damping for the initial data evolution. Simulations are conducted using the Spectral Einstein Code (SpEC) \cite{SXSBBH}. In contrast, the RIT catalog simulations are carried out using the LazEv code \cite{Zlochower:2005bj}, which implements the moving puncture approach \cite{Campanelli:2005dd} along with the BSSNOK (Baumgarte-Shapiro-Shibata-Nakamura-Oohara-Kojima) formalism for evolving systems \cite{Marronetti:2007wz}. The LazEv code interfaces with the Cactus/Carpet/EinsteinToolkit infrastructure \cite{Lousto:2007rj,Zlochower:2012fk,Loffler:2011ay}.

The SXS catalog primarily focuses on circular orbit simulations, while the RIT catalog specializes in eccentric orbit simulations. The fourth release of the RIT catalog comprises 824 eccentric BBH NR simulations, including nonspinning, spin-aligned, and spin-precessing configurations with initial eccentricities ranging from 0 to 1 \cite{Healy:2022wdn}. For our primary research dataset, we select a subset of nonspinning circular orbital simulations from both SXS and RIT, as well as nonspinning eccentric orbits simulations from RIT.

\begin{figure}[htbp!]
\centering
\includegraphics[width=8cm,height=5cm]{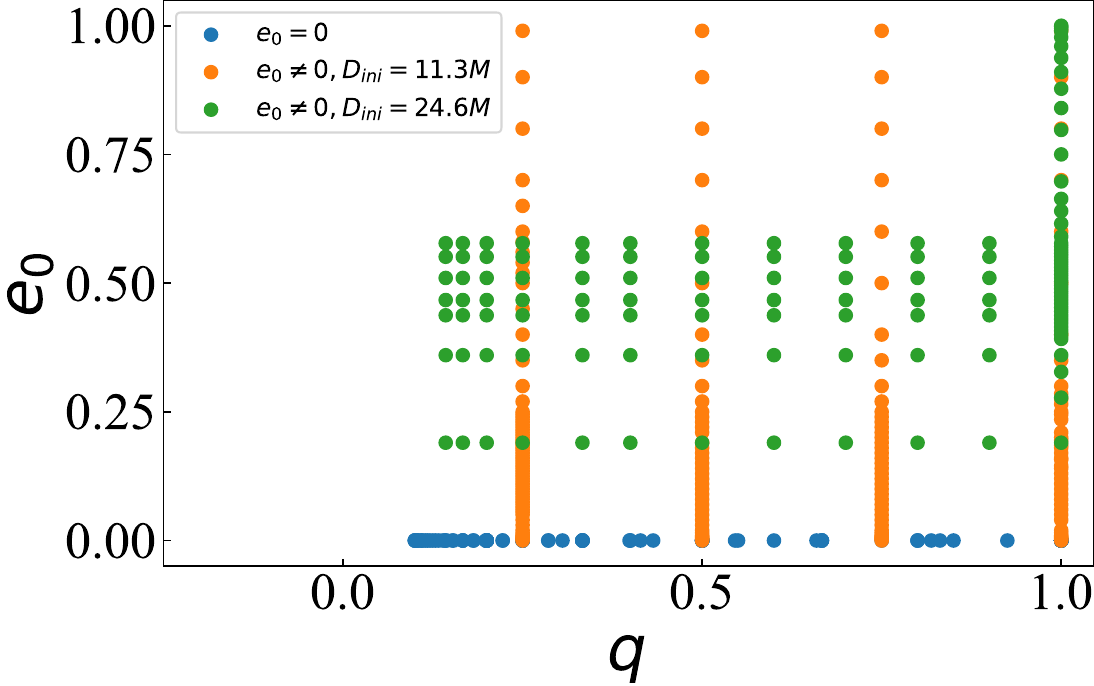}
\caption{\label{FIG:1}Parameter space of all nonspinning eccentric ($e_0\neq0$) orbital and circular ($e_0=0$) orbital simulations from SXS and RIT, which contains 81 sets of SXS circular orbital simulations, 19 sets of RIT circular orbital simulations, and 316 sets of RIT eccentric orbit simulations. The mass ratio $q$ for the circular orbit simulations ranges from 1/10 to 1, while for the eccentric orbit simulations, it spans from 1/7 to 1. The initial eccentricity $e_0$ varies from 0 to 1 for $q = 1,3/4,1/2,1/4$ with initial distances $D_{\text{ini}}=11.3M$ and $q=1$ with $D_{\text{ini}}=24.6M$. For other mass ratios with $D_{\text{ini}}=24.6M$, we only include cases where the orbital cycle number exceeds 1, excluding those with an orbital cycle number less than 1 for the purposes of this study. The calculation of the orbital cycle number can be found in our previous work \cite{Wang:2023vka}.}
\end {figure}

FIG. \ref{FIG:1} illustrates the parameter space encompassing all the eccentric and circular orbital simulation data utilized in this study. RIT generates a continuously varying initial eccentricity by adjusting the tangential linear momentum at two fixed initial distances: $11.3M$ and $24.6M$.
Here, we present 81 sets of SXS circular orbit simulations, 19 sets of RIT circular orbit simulations, and 316 sets of RIT eccentric orbit simulations. 

For the computation of dynamical quantities such as the peak luminosity $L_{\text{peak}}$, remnant mass $M_{\text{rem}}$, spin $\alpha_{\text{rem}}$, and recoil velocity $V_{\text{rem}}$, both SXS and RIT follow similar methodologies. RIT employs AHFinderDirect \cite{Thornburg:2003sf} to identify apparent horizons and determines the horizon spins $S_H/{M_{\text{rem}}}^2$ as the remnant spin $\alpha_{\text{rem}}$ through the isolated horizon algorithm. Subsequently, RIT calculate the horizon mass $m_\text{H}$ as the remnant mass $M_{\text{rem}}$ using the Christodoulou formula: $ m_\text{H} =\sqrt{m_{\mathrm{irr}}^2+S_\text{H}^2 /\left(4 m_{\mathrm{irr}}^2\right)}$, where $m_{\mathrm{irr}} = \sqrt{A_\text{H} /(16 \pi)}$ represents the irreducible mass defined by $A_\text{H}$, the surface area of the horizon \cite{Campanelli:2006fy}. In contrast to mass $M_{\text{rem}}$ and spin $\alpha_{\text{rem}}$, recoil velocity $V_{\text{rem}}$ and peak luminosity $L_{\text{peak}}$ are derived from radiative quantities, i.e., the waveform. 

RIT provides waveform data in the form of the Newman-Penrose scalar $\Psi_4$ and gravitational wave strain $h$, which are accessible from their catalog \cite{RITBBH}. These quantities can be expanded using the spin-weighted spherical harmonic function ${ }_{-2} Y_{\ell , m}(\theta, \phi)$ with a spin weight $s=-2$. Specifically, the expansions are as follows:
\begin{equation}\label{eq:1}
r \Psi_4=\sum_{\ell, m} r \Psi_4^{\ell m}{ }_{-2} Y_{\ell, m}(\theta, \phi),
\end{equation}
and
\begin{equation}\label{eq:2}
r h=\sum_{\ell, m} r h^{\ell m}{}_{-2} Y_{\ell, m}(\theta, \phi),
\end{equation}
where $r$ denotes the extraction radius, and $h^{\ell m}$ and $\Psi_4^{\ell m}$ correspond to higher harmonic modes for $h$ and $\Psi_4$, respectively. As $r$ approaches infinity, the relationship $\Psi_4=\partial^2 h / \partial t^2$ holds. As detailed in Refs. \cite{Ruiz:2007yx}, the radiative quantities can be computed using $h$ and $\Psi_4$. Here we take $h$ as an example.
The radiated energy $E_{\mathrm{rad}}$ can be determined as\cite{Ruiz:2007yx}:
\begin{equation}\label{eq:3}
E_{\mathrm{rad}}=\lim _{r \rightarrow \infty} \frac{r^2}{16 \pi} \sum_{\ell, m} \int_{t_0}^t \mathrm{~d} t^{\prime}\left|\dot{h}^{\ell m}\right|^2,
\end{equation}
where $\dot{h}^{\ell m}$ represents the time derivative of $h^{\ell m}$. Subsequently, the peak luminosity $L_{\text {peak }}$ is calculated as the maximum of $\mathrm{d}E_{\mathrm{rad}}/\mathrm{d}t$:
\begin{equation}\label{eq:4}
L_{\text {peak }}=\max _t \mathrm{d}E_{\mathrm{rad}}/\mathrm{d}t.
\end{equation}
In this study, we focus solely on the nonspinning case, where the recoil velocity $V_{\mathrm{rem}}$ lies in the orbital plane, perpendicular to the orbital angular momentum $L$. It can be calculated as $V_{\mathrm{rem}}=\left|P_{\mathrm{rad}}\right| / M_{\mathrm{rem}}$ with \cite{Ruiz:2007yx}:
\begin{equation}\label{eq:5}
\begin{aligned}
P_{\mathrm{rad}} & =\lim _{r \rightarrow \infty} \frac{r^2}{8 \pi} \sum_{\ell, m} \int_{t_0}^t \mathrm{~d} t^{\prime} \dot{h}^{\ell m} 
\left(a_{\ell, m} \bar{\dot{h}}^{\ell, m+1} \right. \\
&\left. +b_{\ell,-m} \bar{\dot{h}}^{\ell-1, m+1} -b_{\ell+1, m+1} \bar{\dot{h}}^{\ell+1, m+1}\right),
\end{aligned}
\end{equation}
where the coefficients $a_{\ell, m}$, $b_{\ell,-m}$, and $b_{\ell+1, m+1}$ can be referenced from Ref. \cite{Ruiz:2007yx}, and $\bar{\dot{h}}^{\ell m}$ represents the complex conjugate of $\dot{h}^{\ell m}$. Subsequently, we derive the values of both the peak luminosity $L_{\text {peak }}$ and the recoil velocity $V_{\mathrm{rem}}$ from $h$.

\subsection{PN Comparison}\label{sec:II:B}
In Ref. \cite{Wang:2023vka}, we explored the universal oscillation of dynamical quantities $L_{\text{peak}}$, $M_{\text{rem}}$, $\alpha_{\text{rem}}$, and $V_{\text{rem}}$ in relation to initial eccentricity by correlating integer orbital cycle numbers with the peaks and valleys of these oscillations. While this explanation is purely phenomenological, it reveals that the oscillations exhibit a certain periodicity, though it does not uncover the underlying essence of this phenomenon.

In this paper, we aim to analyze the cause of the oscillations from the perspective of waveforms. We focus on the simulation series in which we observed these oscillations in our previous work \cite{Wang:2023vka}, specifically examining five cases with mass ratios $q=1, 3/4,1/2,1/4$ for $D_{\text{ini}}=11.3M$ or $D_{\text{ini}}=24.6M$. Our primary analysis centers on the remnant mass $M_{\text{rem}}$ and spin $\alpha_{\text{rem}}$, while we do not consider the recoil velocity $V_{\text{rem}}$ and peak luminosity $L_{\text{peak}}$. This omission is due to $L_{\text{peak}}$ representing the merger moment, which is dominated by strong field effects. Additionally, according to Ref. \cite{Wang:2023vka}, $V_{\text{rem}}$ varies irregularly with initial eccentricity, complicating the identification of the cause of the oscillations.

The remnant's mass is computed as $M_\text{rem}=M_{\mathrm{ADM}}-E_{\mathrm{rad}}$, where $M_{\mathrm{ADM}}$ denotes the ADM (Arnowitt, Deser, Misner) mass and $E_{\mathrm{rad}}$ represents the gravitational radiative energy. 
We calculate the remnant spin $\alpha_\text{rem}$ using the formula $\alpha_\text{rem}=(L_0-L_z^{\mathrm{rad}})/M_{\text{rem}}^2$, where $L_0$ is the initial orbital angular momentum. Due to symmetry, the radiation of angular momentum $L_{\mathrm{rad}} = L_z^{\mathrm{rad}}$ is concentrated in the $z$ direction (aligned with the direction of the orbital angular momentum $L$). This is calculated by \cite{Ruiz:2007yx} as follows:
\begin{equation}\label{eq:6}
L_{z}^{\mathrm{rad}} =\lim _{r \rightarrow \infty} \frac{ r^2}{16 \pi} \operatorname{Im}\left\{\sum_{\ell, m} m \int_{t_0}^t \left( h^{\ell m} \bar{\dot{h}}^{\ell m} \right)\mathrm{~d} t^{\prime}\right\}.
\end{equation}

\begin{figure*}[htbp!]
\centering
\includegraphics[width=16cm,height=5cm]{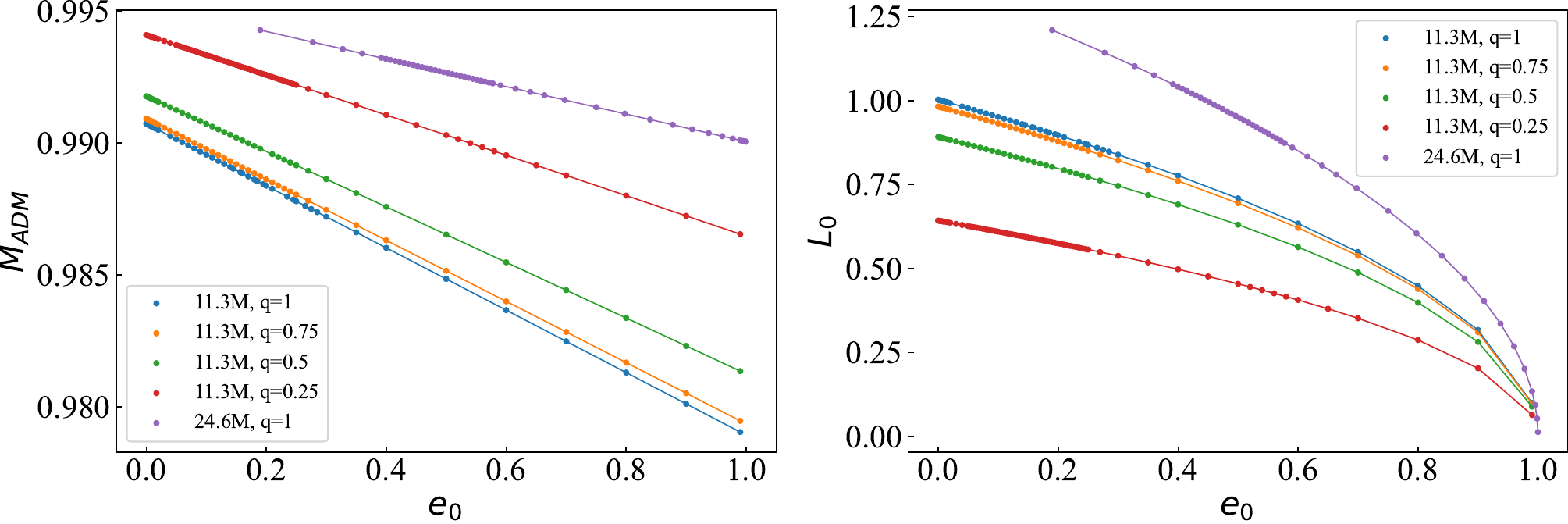}
\caption{\label{FIG:2}
Relationship between initial ADM mass $M_{\mathrm{ADM}}$ and angular momentum $L_0$ with initial eccentricity $e_0$ in the RIT simulation series $q=1, 3/4,1/2,1/4$ for $D_{\text{ini}}=11.3M$ and $q=1$ for $D_{\text{ini}}=24.6M$. Each point represents a NR simulation, and the lines of the same color are drawn just to highlight the trend of change.}
\end {figure*}

To investigate the oscillation of $M_\text{rem}$ and $\alpha_\text{rem}$ with initial eccentricity $e_0$, we first analyze the relationship between $M_{\mathrm{ADM}}$ and $L_0$ as a function of $e_0$. In FIG. \ref{FIG:2}, we present this relationship for the RIT simulation series with mass ratios $q=1, 3/4,1/2,1/4$ at $D_{\text{ini}}=11.3M$ and for $q=1$ at $D_{\text{ini}}=24.6M$. From FIG. \ref{FIG:2}, it is evident that both $M_{\mathrm{ADM}}$ and $L_0$ vary smoothly with $e_0$ without exhibiting oscillations. Thus, the oscillation is not attributed to $M_{\mathrm{ADM}}$ and $L_0$. 

\begin{figure*}[htbp!]
\centering
\includegraphics[width=16cm,height=5cm]{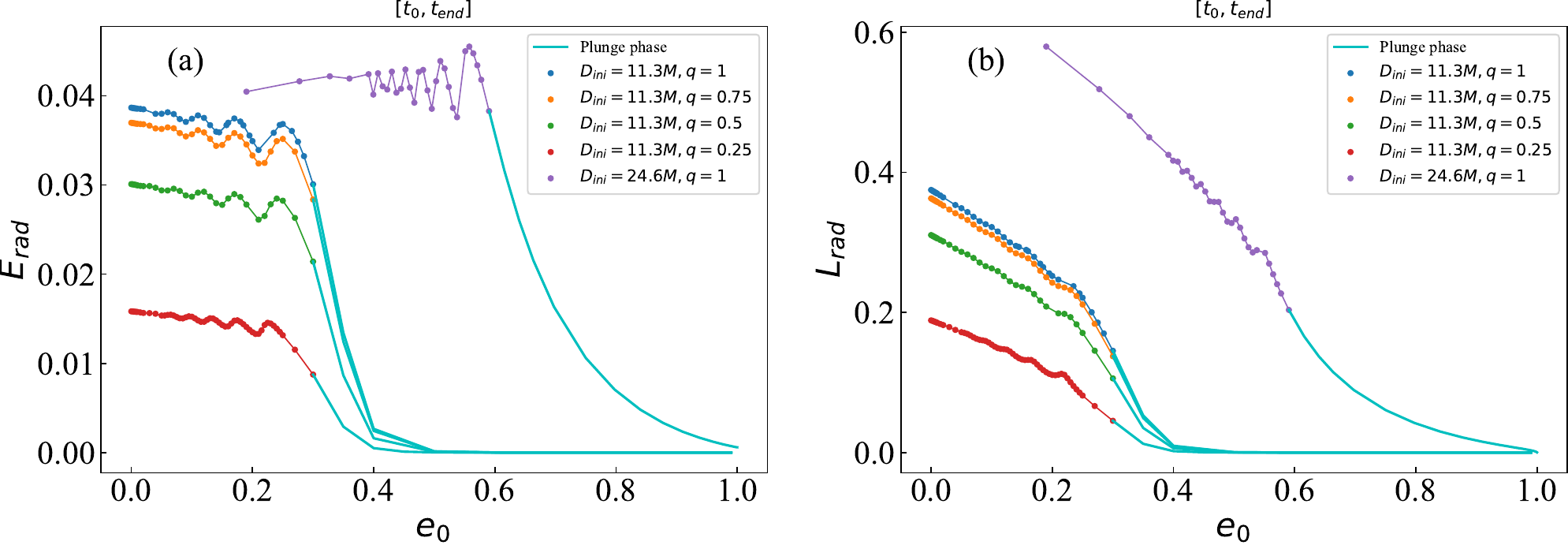}
\caption{\label{FIG:3}Variation of radiated energy $E_{\mathrm{rad}}$ and radiated angular momentum $L_{\mathrm{rad}}$ with the initial eccentricity $e_0$ for all oscillating waveforms, which are calculated by $h$ from the initial time $t_0$ to the end time $t_{\text{end}}$. The blue solid line in the figures indicates BBH orbital motion lasting less than one cycle, corresponding to the plunge phase, during which no oscillatory behavior is observed; consequently, this phase is not shown explicitly.}
\end {figure*}

Next, we turn our attention to $E_{\mathrm{rad}}$ and $L_{\mathrm{rad}}$, which are likely responsible for the observed oscillations. In FIG. \ref{FIG:3}, we illustrate the variations of $E_{\mathrm{rad}}$ and $L_{\mathrm{rad}}$ with respect to $e_0$ for all corresponding oscillating waveforms. These quantities are calculated from $h$ over the time interval from initial time $t_0$ to end time $t_{\text{end}}$ using Eqs. (\ref{eq:3}) and (\ref{eq:6}). It is evident that both $E_{\mathrm{rad}}$ and $L_{\mathrm{rad}}$ exhibit oscillations as $e_0$ changes, and this oscillation pattern is consistent with those observed in $M_\text{rem}$ and $\alpha_\text{rem}$. Therefore, we conclude that the oscillations originate from the waveforms or the radiative quantities $E_{\mathrm{rad}}$ and $L_{\mathrm{rad}}$. 

It is noteworthy that there are differences between the oscillations of $E_{\mathrm{rad}}$ and $L_{\mathrm{rad}}$ compared to those of $M_\text{rem}$ and $\alpha_\text{rem}$ with eccentricity. Specifically, the oscillations of $E_{\mathrm{rad}}$ and $L_{\mathrm{rad}}$ exhibit a slow decline, whereas the oscillations of $M_\text{rem}$ and $\alpha_\text{rem}$ appear nearly horizontal. This discrepancy is attributed to the gradual decline of $M_{\mathrm{ADM}}$ and $L_0$. In FIG. \ref{FIG:3}, the blue solid line indicates that the orbital motion of the BBH is less than one orbital cycle, corresponding to the plunge phase, during which no oscillation behavior is observed. Consequently, this phase is not the focus of our analysis.

\begin{figure*}[htbp!]
\centering
\includegraphics[width=16cm,height=10cm]{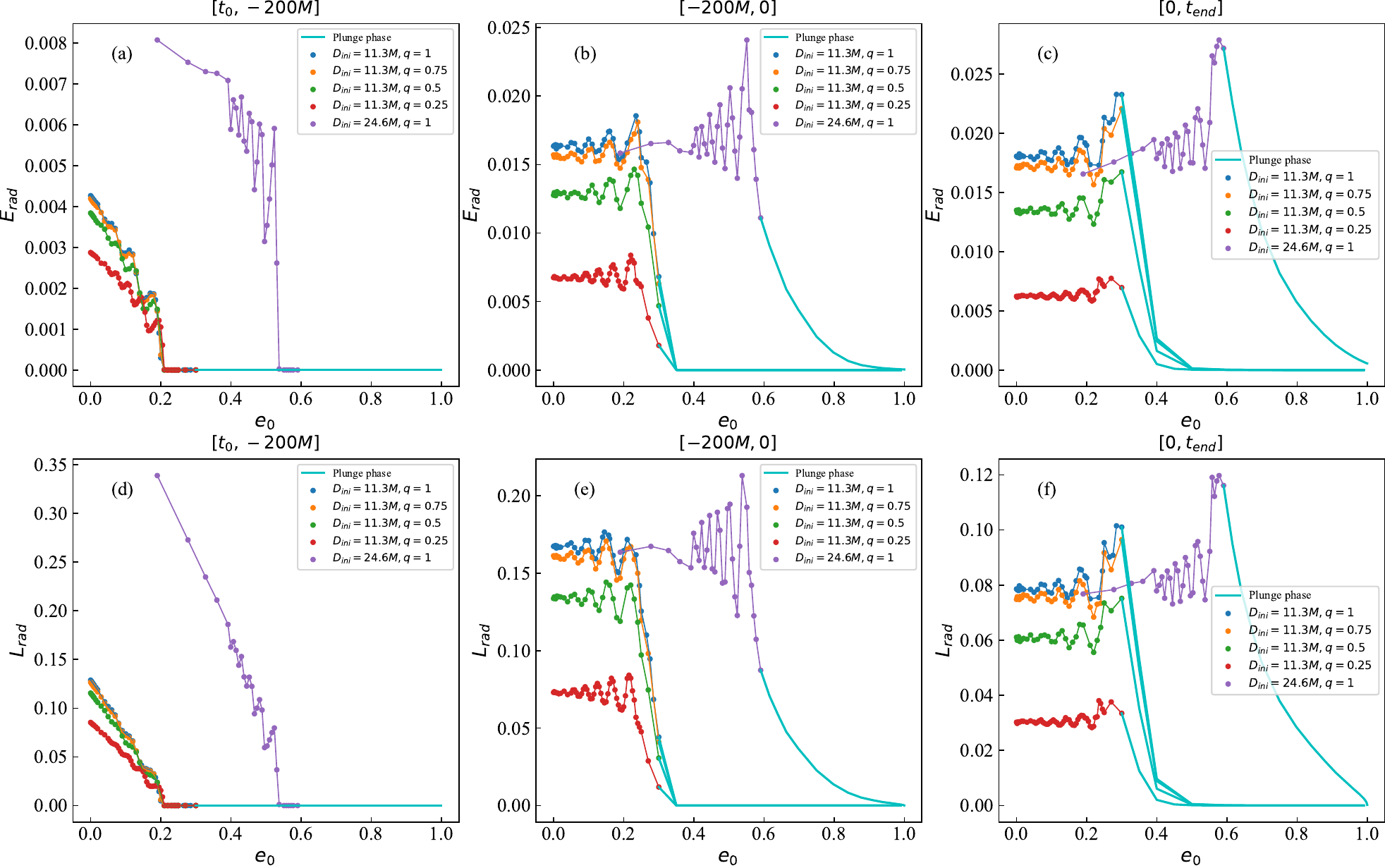}
\caption{\label{FIG:4}
Relationship between the radiative energy $E_{\mathrm{rad}}$ and angular momentum $L_{\mathrm{rad}}$ and the initial eccentricity $e_0$ in three phases $[t_0,-200M]$ ((a), (d)), $[-200M,0]$ ((b), (e)), $[0,t_{\text{end}}]$ ((c), (f)). Similar to the FIG. \ref{FIG:3}, the blue solid line in the figures indicates BBH orbital motion lasting less than one cycle, corresponding to the plunge phase, during which no oscillatory behavior is observed; consequently, this phase is not shown explicitly.}
\end {figure*}

In order to analyze the cause of the oscillations, we decompose the calculation of the radiative quantities $E_{\mathrm{rad}}$ and $L_{\mathrm{rad}}$ into three phases based on waveform time: inspiral, late inspiral to merger, and merger to ringdown. The inspiral phase spans from the initial moment $t_0$ to $-200M$ (with $200M$ preceding the merger, where we designate the merger time as 0); the late inspiral to merger phase covers from $-200M$ to 0; and the subsequent ringdown phase commences beyond 0.
In FIG. \ref{FIG:4}, we show the relationship between the radiative energy $E_{\mathrm{rad}}$ and angular momentum $L_{\mathrm{rad}}$ and the initial eccentricity $e_0$ across the three phases $[t_0,-200M]$, $[-200M,0]$, $[0,t_{\text{end}}]$.  
The figure demonstrates that both $E_{\mathrm{rad}}$ and $L_{\mathrm{rad}}$ oscillate with initial eccentricity in all three phases, and their oscillation modes are consistent, indicating that the underlying causes of these oscillations are the same.
In panels (a) and (d), some values of $E_{\mathrm{rad}}$ and $L_{\mathrm{rad}}$ are zero due to the waveform duration being less than $200M$. From FIG. \ref{FIG:4}, we conclude that oscillations are present in the inspiral, late inspiral to merger, and merger to ringdown phases. Based on the oscillation of peak luminosity and amplitudes discussed in Ref. \cite{Wang:2023wol}, we know that these oscillations also occur at the moment of merger. Thus, the influence of oscillations is evident in all phases of eccentric BBH mergers, whereas such oscillations are absent in circular orbital BBH mergers, as noted in Ref. \cite{Wang:2023vka}.
It is important to emphasize that the reference time does not have to be restricted to $-200M$; similar results are observed at $-300M$, where oscillations also persist.

\begin{figure*}[htbp!]
\centering
\includegraphics[width=16cm,height=5cm]{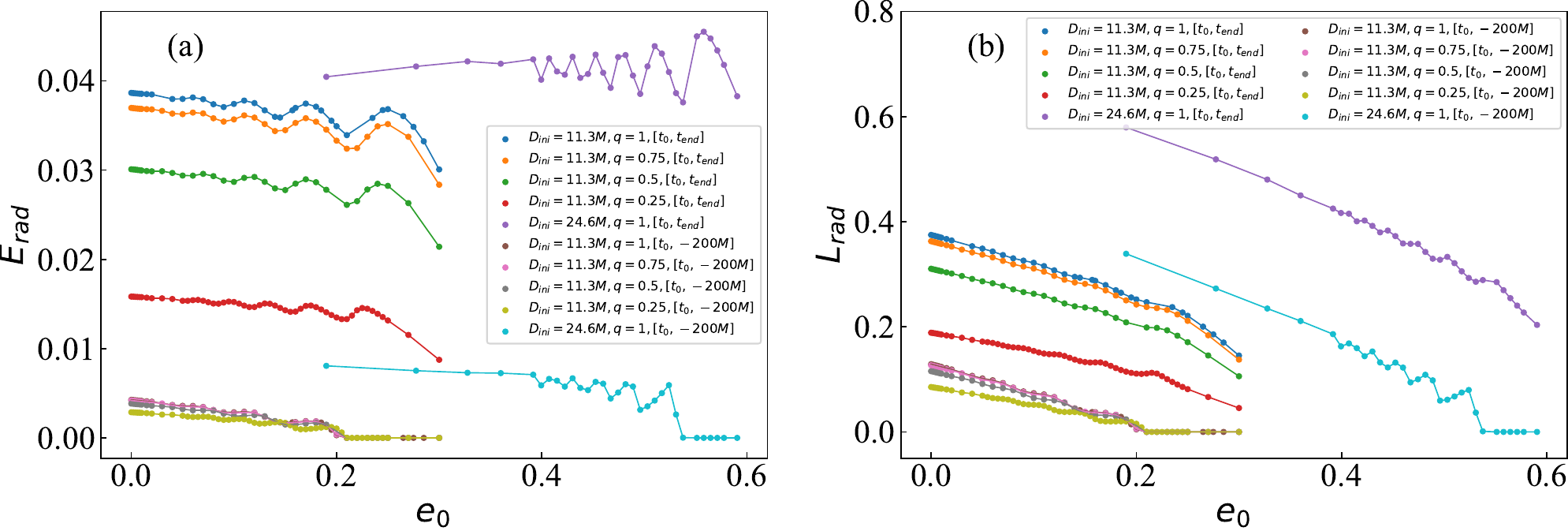}
\caption{\label{FIG:5}Comparison of radiative energy $E_{\mathrm{rad}}$ and angular momentum $L_{\mathrm{rad}}$ between panels (a) and (d) in FIG. \ref{FIG:4} and FIG. \ref{FIG:3}. The former corresponds to the inspiral phase of the waveforms, while the latter covers the entire duration; nevertheless, both display the same oscillatory modes.}
\end {figure*}

To facilitate a more intuitive comparison, FIG. \ref{FIG:5} presents a comparison of $E_{\mathrm{rad}}$ and $L_{\mathrm{rad}}$ between panels (a) and (d) in FIG. \ref{FIG:4} and FIG. \ref{FIG:3}. From FIG. \ref{FIG:5}, it is evident that the two oscillation modes are consistent, indicating that they share the same physical origin.
The late inspiral to merger and merger to ringdown phases are characterized by strong field effects in the dynamics of BBHs, which are challenging to analyze analytically. Therefore, we concentrate on the inspiral phase, where analytical PN methods can provide valuable insights.

In Ref. \cite{Wang:2024jro}, we conducted a comprehensive comparison of PN and NR waveforms for eccentric BBH orbits. This comparison encompasses 180 sets of nonspinning and spin-aligned eccentric waveforms from RIT and SXS, including various gravitational wave harmonic modes such as $(\ell , m)$ = (2, 2), (2, 1), (3, 3), (3, 2), (4, 4) and (5, 5). We employed the 3PN quasi-Kepler parameterization, 3PN radiative backaction, and 3PN high-order moments to generate the eccentric PN waveforms.
Our findings indicate that the frequency and amplitude of the PN and NR waveforms are highly consistent up to $200M$ before merger, demonstrating that the PN waveform accurately reproduces the NR waveform prior to $200M$ when appropriate initial parameters are used. 
According to Ref. \cite{Wang:2024jro}, there are four initial parameters required to generate a complete eccentric PN waveform: the PN expansion parameter $x_0$ defined as $x \equiv(M \omega)^{2 / 3}$ (where $\omega$ is the average orbital frequency), the time eccentricity ${e_t}_0$, the mean anomaly $l_0$, and the initial phase $\varphi_0$. Since $\varphi_0$ is a free parameter, only three real parameters $x_0$, ${e_t}_0$, and $l_0$ are needed to generate an eccentric PN waveform.
Among these initial parameters, ${e_t}_0$ is a focus of our research, contributing to the gradual increase of oscillations observed in FIG. \ref{FIG:3}, though it is not the source of the oscillation itself. The parameter $x$ dictates the initial frequency of the waveform and influences its duration. Our analysis in Ref. \cite{Wang:2024jro} reveals that the effect of $\omega$ in $x$ closely resembles the frequency of the corresponding circular orbital waveform, indicating that $x_0$ is not responsible for the oscillation. Therefore, we conclude that the remaining parameter, $l_0$, is likely the cause of the observed oscillations.

\begin{figure*}[htbp!]
\centering
\includegraphics[width=16cm,height=20cm]{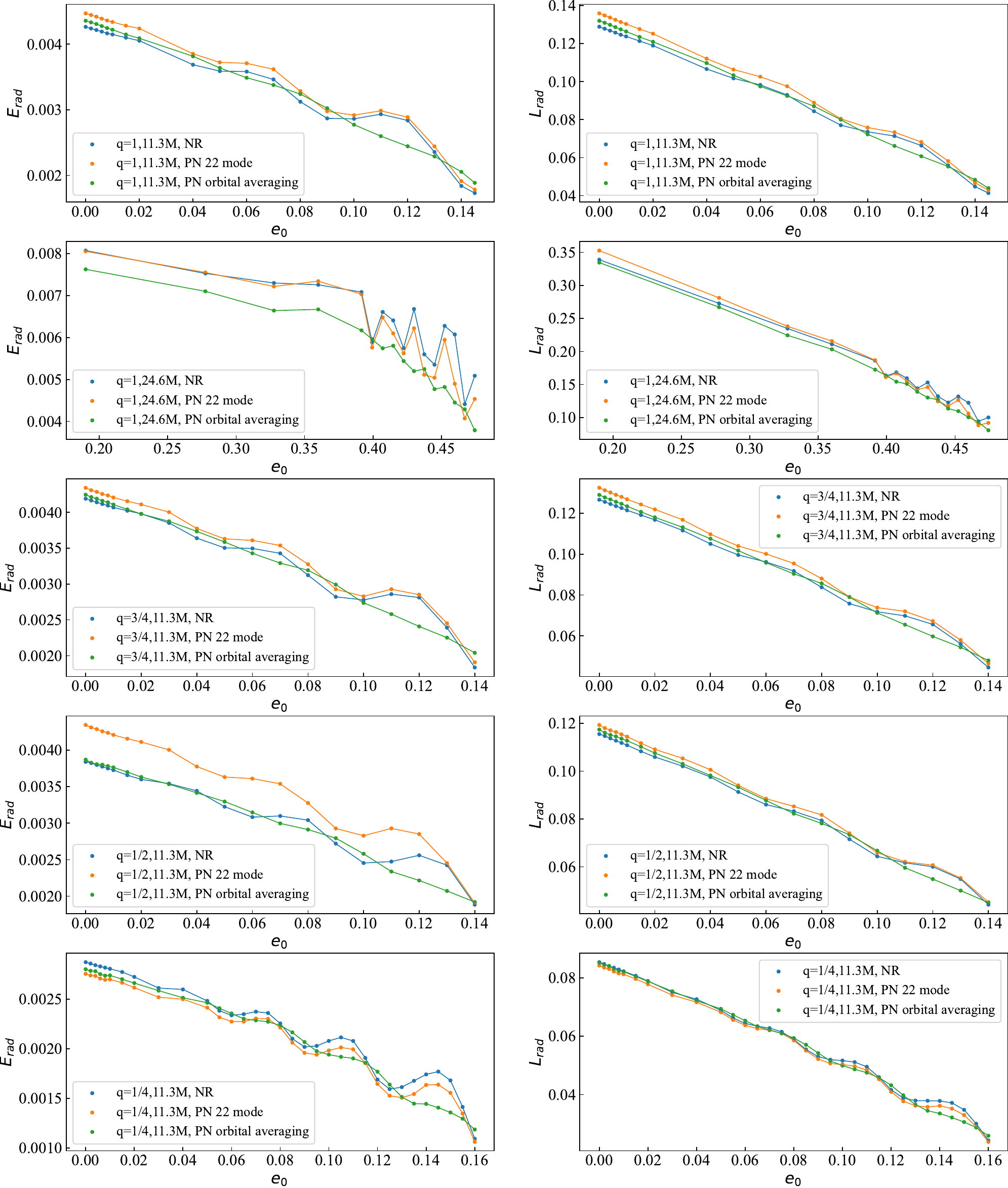}
\caption{\label{FIG:6}NR radiated energy and angular momentum corresponding to the inspiral phase in panels (a) and (d) of FIG. \ref{FIG:3} (labeled by `NR'), as well as the radiated energy and angular momentum calculated using the 3PN waveform with initial parameters fitting to the NR waveform (labeled by `PN 22 mode'),  and the orbital averaged radiated energy and angular momentum with the same initial parameters (labeled by `PN orbital averaging').}
\end {figure*}

According to Ref. \cite{Wang:2024jro}, PN methods can accurately produce waveforms in the range $[t_0, -200M]$, specifically during the inspiral phase. Therefore, we focus on panels (a) and (d) in FIG. \ref{FIG:4}, which represent this inspiral phase, to calculate the radiated energy and angular momentum, aiming to identify the origin of the oscillations using PN waveforms.
To investigate the effect of $l_0$, we calculate the energy and angular momentum and orbital averaged energy and angular momentum by 3PN (2,2) modes of waveform and 3PN orbital average energy flux and angular momentum flux using the same initial PN parameters as the NR waveform by PN fitting to NR in Ref. \cite{Wang:2024jro}, and compare them with the energy flux and angular momentum flux calculated by NR waveforms. Radiated energy and angular momentum for both the NR and 3PN (2,2) modes are computed using Eqs. (\ref{eq:3}) and (\ref{eq:6}). 

The orbital average of the 3PN energy and angular momentum fluxes includes both instantaneous and hereditary contributions, such as tail effects, tail-of-tails, and non-linear memory \cite{Arun:2009mc}, which drive the time evolution of $x$ and $e_t$. The orbital average functions as a periodic average, designed to eliminate the influence of the mean anomaly $l_0$. According to Ref. \cite{Arun:2009mc}, the orbital average energy flux and angular momentum flux in 3PN formalism, expressed in ADM coordinates (which can be transformed into harmonic coordinates via a coordinate transformation), can be represented as the sum of the instantaneous and hereditary terms:
\begin{equation}\label{eq:7}
\langle\mathcal{F}\rangle=\left\langle\mathcal{F}_{\text {inst }}\right\rangle+\left\langle\mathcal{F}_{\text {hered }}\right\rangle,
\end{equation}
\begin{equation}\label{eq:8}
\langle\mathcal{G}\rangle=\left\langle\mathcal{G}_{\text {inst }}\right\rangle+\left\langle\mathcal{G}_{\text {hered }}\right\rangle,
\end{equation}
where $\langle\mathcal{F}\rangle$ and $\langle\mathcal{G}\rangle$ represents the orbital average or periodic average of $\mathcal{F}$ and $\mathcal{G}$, and the instantaneous term and hereditary term are expressed as:
\begin{widetext}
\begin{equation}
\begin{aligned}
\left\langle\mathcal{F}_{\text {inst }}\right\rangle &=\frac{32}{5} \frac{c^5}{G} \eta^2 x^5\left(\mathcal{F}_{\text {Newt }}+\mathcal{F}_{1 \mathrm{PN}} x+\mathcal{F}_{2 \mathrm{PN}} x^2+\mathcal{F}_{3 \mathrm{PN}} x^3\right), \\
\left\langle\mathcal{F}_{\text {hered }}\right\rangle&=\frac{32}{5} \frac{c^5}{G} \eta^2 x^5\left(\mathcal{F}_{1.5 \mathrm{PN}} x^{3 / 2}+\mathcal{F}_{2.5 \mathrm{PN}} x^{5 / 2}+\mathcal{F}_{3 \mathrm{PN}} x^3\right),
\end{aligned}
\end{equation}
\begin{equation}
\begin{aligned}
\left\langle\mathcal{G}_{\text {inst }}\right\rangle&=\frac{4}{5} c^2 m \eta^2 x^{7 / 2}\left(\mathcal{G}_{\mathrm{N}}+x \mathcal{G}_{1 \mathrm{PN}}+x^2 \mathcal{G}_{2 \mathrm{PN}}+x^3 \mathcal{G}_{3 \mathrm{PN}}\right),\\
\left\langle\mathcal{G}_{\text {hered }}\right\rangle&=\frac{32}{5} {c^2}m \eta^2 x^{7/2}\left(\mathcal{G}_{1.5 \mathrm{PN}} x^{3 / 2}+\mathcal{G}_{2.5 \mathrm{PN}} x^{5 / 2}+\mathcal{G}_{3 \mathrm{PN}} x^3\right),
\end{aligned}
\end{equation}
\end{widetext}
where $\mathcal{F}_{\text {Newt }}$, $\mathcal{G}_{\text {Newt }}$... represents the PN expansion coefficient, and its specific expression can be found in Appendix \ref{App:A}. 

We calculate the orbital averaged energy and angular momentum by integrating $\left\langle\mathcal{F}_{\text {inst }}\right\rangle=\mathrm{d} E / \mathrm{d} t$ and $\left\langle\mathcal{G}_{\text {inst }}\right\rangle=\mathrm{d} L / \mathrm{d} t$ as given in Eqs. (\ref{eq:7}) and (\ref{eq:8}). The initial parameters used in these calculations are $x_0$ and ${e_t}_0$, which are obtained from PN fitting to the NR waveform.

In FIG. \ref{FIG:6}, we present the NR radiated energy and angular momentum corresponding to the inspiral phase shown in panels (a) and (d) of FIG. \ref{FIG:3}. Additionally, we include the radiated energy and angular momentum calculated using the 3PN waveform with initial parameters fitted to the NR waveform, as well as the orbital averaged radiated energy and angular momentum using the same initial parameters. For the PN calculations, we focus solely on the (2,2) mode, as its contribution is the most significant. Furthermore, Ref. \cite{Wang:2024jro} indicates that including other modes does not enhance the accuracy of the results.
From FIG. \ref{FIG:6}, it is evident that the radiated energy and angular momentum calculated for both the NR and PN (2,2) modes exhibit similar oscillation patterns; however, the PN orbital averaged energy and angular momentum do not display oscillations. The oscillation mode and its orbital average for the case of $24.6M$ in FIG. \ref{FIG:6} are less pronounced than those at $11.3M$, primarily due to the smaller data set for the former compared to the latter. 
We ensure consistency in our representation of eccentricity by utilizing the initial eccentricity $e_0$ provided by RIT, along with the time eccentricity ${e_t}_0$ from PN theory. As noted in Ref. \cite{Wang:2024jro}, the difference between these two measures is minimal.
The radiated energy of the PN (2,2) mode shows some deviation from that of the NR, attributed to a larger amplitude error between PN and NR compared to the frequency error, as indicated by the findings in Ref. \cite{Wang:2024jro}. This discrepancy causes the orbital average in FIG. \ref{FIG:6} to align closely with NR, while the results of the PN (2,2) mode deviate.
The absence of oscillation in the orbital average, alongside the contrasting oscillations in the NR and PN (2,2) modes depicted in FIG. \ref{FIG:6}, underscores the influence of the mean anomaly $l_0$ on the radiated energy and angular momentum. This effect is smoothed out during the calculation of the orbital averaged energy flux and angular momentum flux. The presence of $l_0$ signifies an orbital influence that shapes the entire process of BBH merger in eccentric orbits, as illustrated in FIG. \ref{FIG:4}.
 
\begin{figure*}[htbp!]
\centering
\includegraphics[width=16cm,height=8cm]{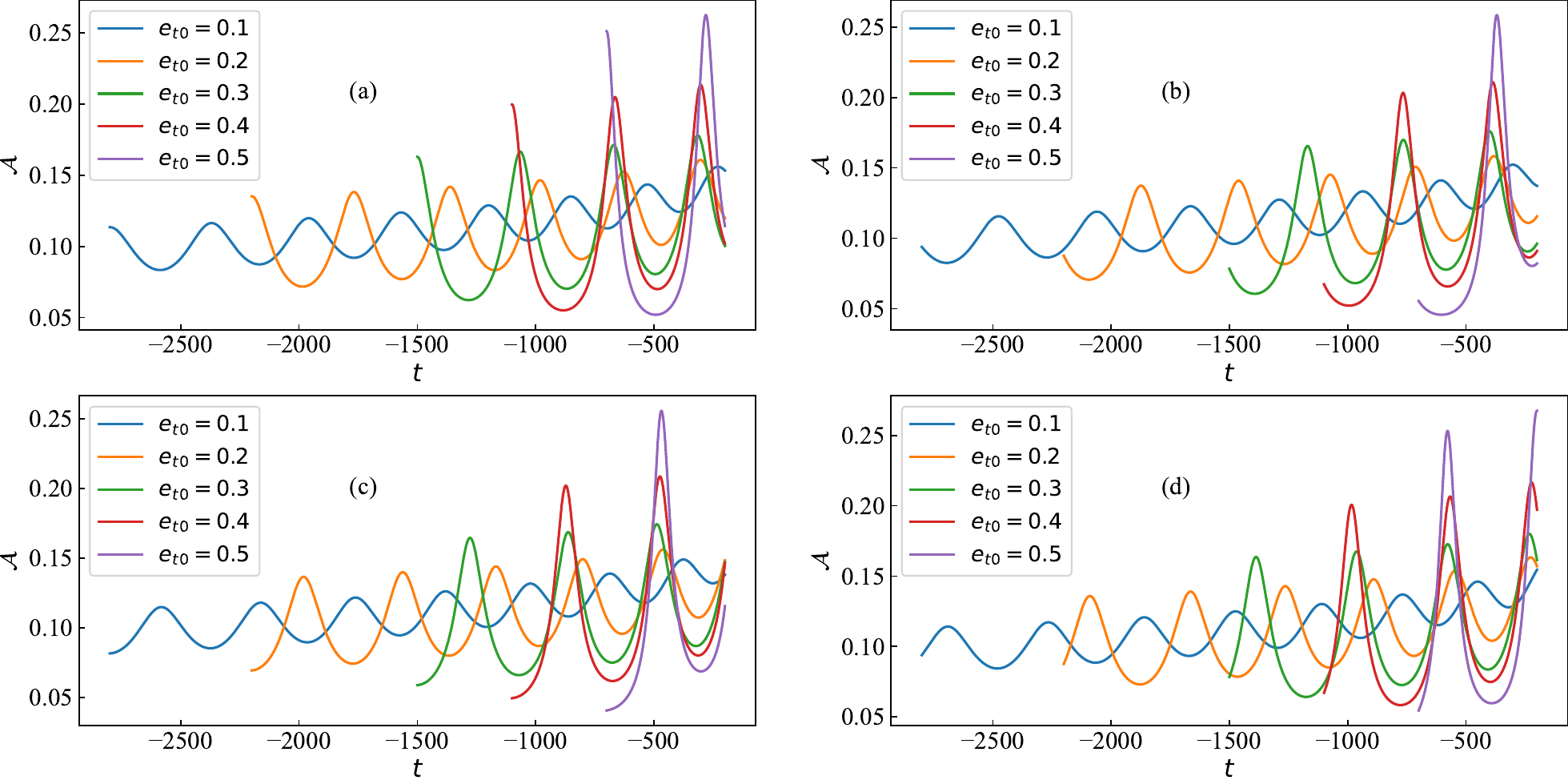}
\caption{\label{FIG:7}Five sets of amplitude of waveforms with initial eccentricities increasing from ${e_t}_0=0.1$ to ${e_t}_0=0.5$ and $q=1/2$. Their $x_0=0.07$ and cutoff time $t=-200M$ are the same, but their mean anomalies from panel (a) to (d) are 0, $\pi/2$, $\pi$, and $3\pi/2$ respectively. The waveform corresponding to $l_0=2\pi$ is identical to that of $l_0=0$ and is therefore not shown.}
\end {figure*}

We now examine the impact of the initial mean anomaly $l_0$ on the waveform from a comprehensive PN perspective. In FIG. \ref{FIG:7}, we present five sets of waveform amplitudes with initial eccentricities ranging from ${e_t}_0=0.1$ to ${e_t}_0=0.5$ for a mass ratio of $q=1/2$. All waveforms share the same initial parameter $x_0=0.07$ and cutoff time $t=-200M$, while their mean anomalies in panels (a) to (d) are 0, $\pi/2$, $\pi$, and $3\pi/2$, respectively. The waveform corresponding to $l_0=2\pi$ is identical to that of $l_0=0$ and is therefore not shown.

From FIG. \ref{FIG:7}, it is evident that both the initial eccentricity and the initial mean anomaly significantly influence the amplitude of the waveform. We focus primarily on the amplitude because the radiated energy and angular momentum calculated using Eqs. (\ref{eq:3}) and (\ref{eq:6}) are primarily related to amplitude rather than frequency and phase. The initial eccentricity determines the magnitude of the amplitude oscillation, while the initial mean anomaly specifies the initial position of the waveform's amplitude, which may correspond to the pericenter ($l_0 = 0$), the apocenter ($l_0 = \pi$), or other positions. Notably, for waveforms with the same initial eccentricity, their orbital averages of amplitudes remain consistent, although their initial mean anomalies are different. This behavior is analogous to the fact that the orbital average of energy and angular momentum is independent of $l_0$.

\begin{figure*}[htbp!]
\centering
\includegraphics[width=16cm,height=5cm]{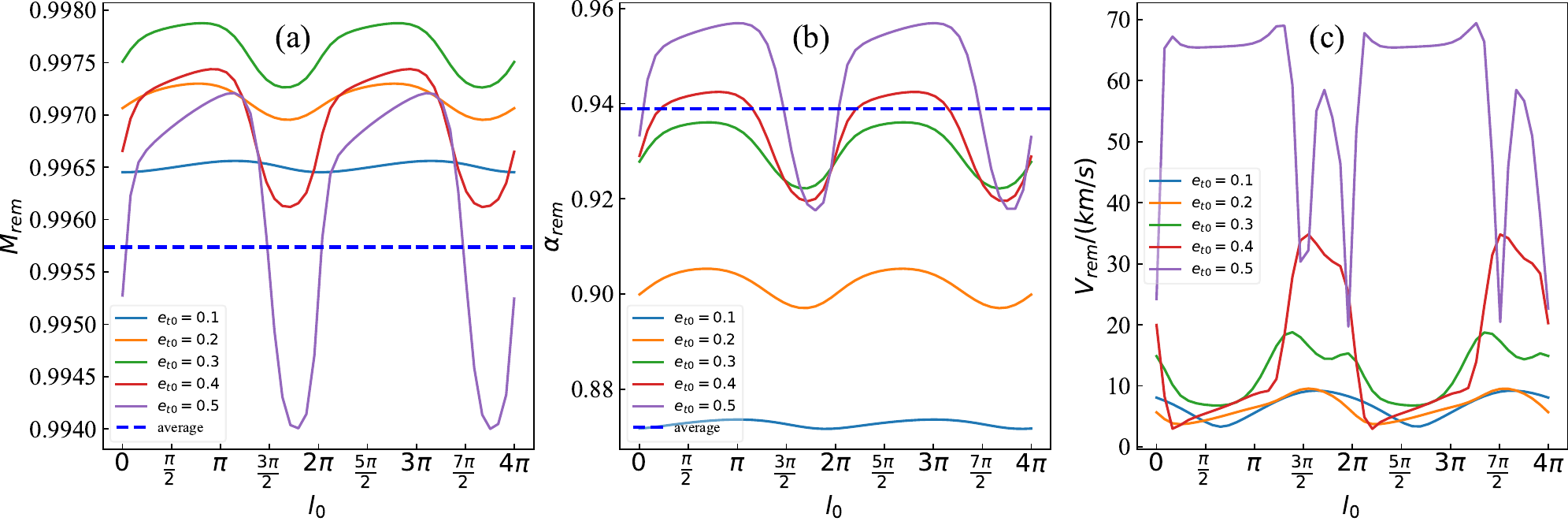}
\caption{\label{FIG:8}$M_\text{rem}$, $\alpha_\text{rem}$, and $V_\text{rem}$ calculated corresponding to the waveforms of FIG. \ref{FIG:7} traversing the parameter space $[0,2\pi]$ of the initial mean anomaly $l_0$. The solid line shows the dynamical quantities obtained by traversing 
$l_0$, while the dotted line represents the PN orbital average for 
${e_t}_0 = 0.5$.}
\end {figure*}

We utilize the PN waveforms shown in FIG. \ref{FIG:7} to explore the parameter space $[0,2\pi]$ of the initial mean anomaly $l_0$ and calculate the mass, spin, and recoil velocity of the remnant black hole in order to observe the effects of varying $l_0$. In our calculations, we assume that radiation during the merger and ringdown phases, occurring after $t=-200M$, is negligible, as these phases are beyond the scope of PN theory and do not contribute to the elucidation of the problem.
And we assume $M_{\mathrm{ADM}}=1$, neglecting the effect of binding energy $E_{\mathrm{b}}=M_{\mathrm{ADM}}-M$, whose contribution does not affect our final results.. Additionally, we set $L_0=1$ to simplify the calculations.
In FIG. \ref{FIG:8}, we present the results for the remnant mass $M_\text{rem}$, remnant spin $\alpha_\text{rem}$, and recoil velocity $V_\text{rem}$ as we traverse the parameter space $[0,2\pi]$ of the initial mean anomaly $l_0$. The results corresponding to the initial eccentricity of 0.5 are highlighted for the orbital average calculation. While we can compute the orbital averages for other cases, those results are not marked for clarity.
In FIG. \ref{FIG:8}, both $M_\text{rem}$ and $\alpha_\text{rem}$ are calculated using the (2,2) mode, while $V_\text{rem}$ is derived from the $\ell \leq 4$ modes, as higher harmonic modes significantly influence the recoil velocity, according to Ref. \cite{Radia:2021hjs}. The average recoil velocity was not calculated due to the lack of corresponding calculation formula.

From FIG. \ref{FIG:8}, it is evident that for a constant initial eccentricity, varying $l_0$ induces oscillations in these dynamical quantities; notably, higher initial eccentricities result in more pronounced oscillations. The figure clearly illustrates the increasing amplitude of oscillations in dynamical quantities as eccentricity rises, consistent with the observations in FIG. \ref{FIG:3}. Specifically, for mass, spin, and recoil velocity, the amplitude of oscillations at ${e_t}_0 = 0.5$ is enhanced by approximately 10, 5, and 7 times, respectively, compared to those at ${e_t}_0 = 0.1$. It is anticipated that in a circular orbit, i.e., for ${e_t}_0 = 0$, the oscillations will vanish.
It is important to note that the average values of these oscillations do not align perfectly, as we have not accounted for the merger and ringdown phases. Including these phases would yield average values that are more closely matched, as demonstrated in Ref. \cite{Wang:2023vka}. Additionally, the recoil oscillations shown in panel (c) exhibit less regularity compared to those in panels (a) and (b). This irregularity arises from the recoil calculation formula (Eq. (\ref{eq:5})), which incorporates cross terms of harmonic modes, such as the interaction between the (2,2) and (3,3) modes. This interaction fundamentally contributes to the recoil anomalies discussed in Ref. \cite{Radia:2021hjs}. FIG. \ref{FIG:8} illustrates the impact of the mean anomaly during the inspiral phase; this effect would be significantly stronger if the merger and ringdown phases were also considered, as indicated in panels (b), (c), (e), and (f) of FIG. \ref{FIG:4}.

FIG. \ref{FIG:3} demonstrates that the effect of mean anomaly is present throughout all phases, particularly in cases of high eccentricity. As eccentricity increases, the oscillations become more pronounced. When only one complete orbit remains, this oscillation reaches its maximum magnitude, as noted in Ref. \cite{Wang:2023vka}, coinciding with the maximum of the initial eccentricity for a certain initial distance. This observation is consistent with the principles illustrated in FIG. \ref{FIG:8}. This oscillatory behavior encapsulates how the mean anomaly influences the dynamics of eccentric BBH mergers, particularly regarding peak amplitude of gravitational waveform, which reflects the full impact of strong field effects.

However, the initial distance that NR can simulate is limited, constraining the peak values of dynamical quantities observed at medium eccentricities (e.g., $e_0 = 0.3$ for $11.3M$ and $e_0 = 0.6$ for $24.6M$). In reality, higher initial eccentricities would yield even greater enhancements, but achieving this necessitates a larger initial distance for NR simulations.

\begin{figure*}[htbp!]
\centering
\includegraphics[width=16cm,height=5cm]{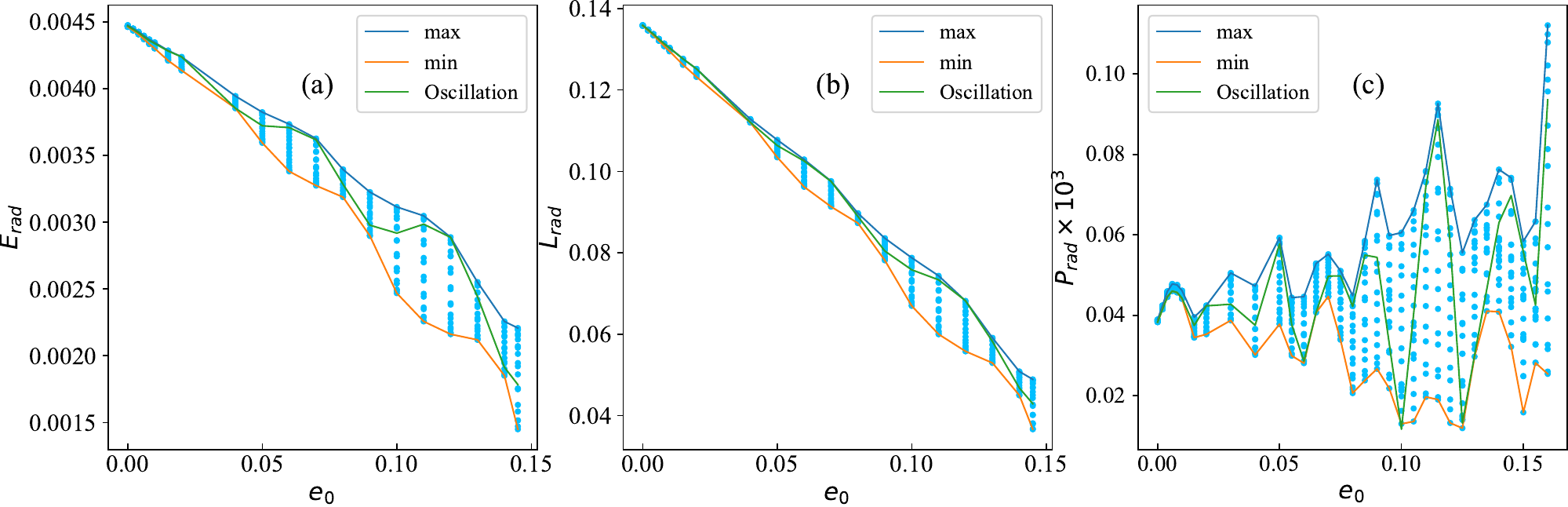}
\caption{\label{FIG:9}Values and ranges of radiative quantities—radiated energy $E_{\mathrm{rad}}$ (a), radiated angular momentum $L_{\mathrm{rad}}$ (b), and radiated linear momentum $P_{\mathrm{rad}}$ (c), which are calculated from PN waveforms fitted to NR waveforms during the inspiral phase ($200M$ before the merger) for specific mass ratios, as the initial mean anomaly $l_0$ varies continuously from 0 to $2\pi$. The broken lines marked with `max’ and `min’ in the figures represent the maximum and minimum values of the same eccentricity. The scattered points in the middle of the lines represent the values of the radiative quantities generated after traversing $l_0$. The green broken lines marked as `oscillation' are the specific radiative quantities of the corresponding inspiral part in FIG.  \ref{FIG:4}.}
\end {figure*}

\begin{figure*}[htbp!]
\centering
\includegraphics[width=16cm,height=15cm]{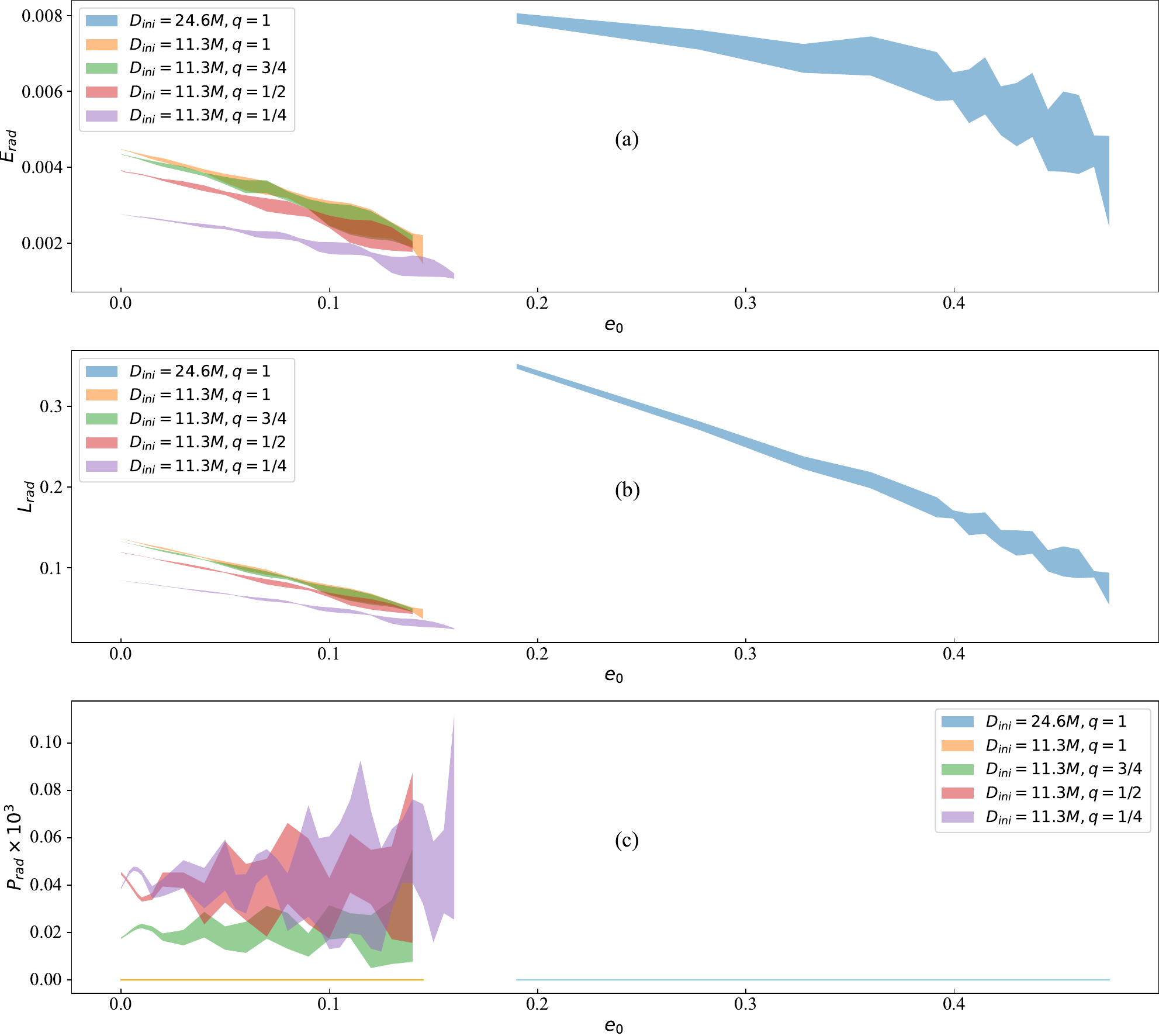}
\caption{\label{FIG:10}Envelopes of radiated energy $E_{\mathrm{rad}}$ (a), radiated angular momentum $L_{\mathrm{rad}}$ (b), and radiated linear momentum $P_{\mathrm{rad}}$ (c) delineated by the five cases ($D_{\text{ini}}=24.6 M$ with $q=1$, $D_{\text{ini}}=11.3 M$ with $q=1$, $D_{\text{ini}}=11.3 M$ with $q=3/4$, $D_{\text{ini}}=11.3 M$ with $q=1/2$, and $D_{\text{ini}}=11.3 M$ with $q=1/4$) with the initial eccentricity for varying $l_0$ in $[0,2\pi]$. Given that linear momentum radiation is null for mass ratio of $q=1$, it is represented solely as a line in panel (c).}
\end {figure*}

\begin{figure*}[htbp!]
\centering
\includegraphics[width=16cm,height=8cm]{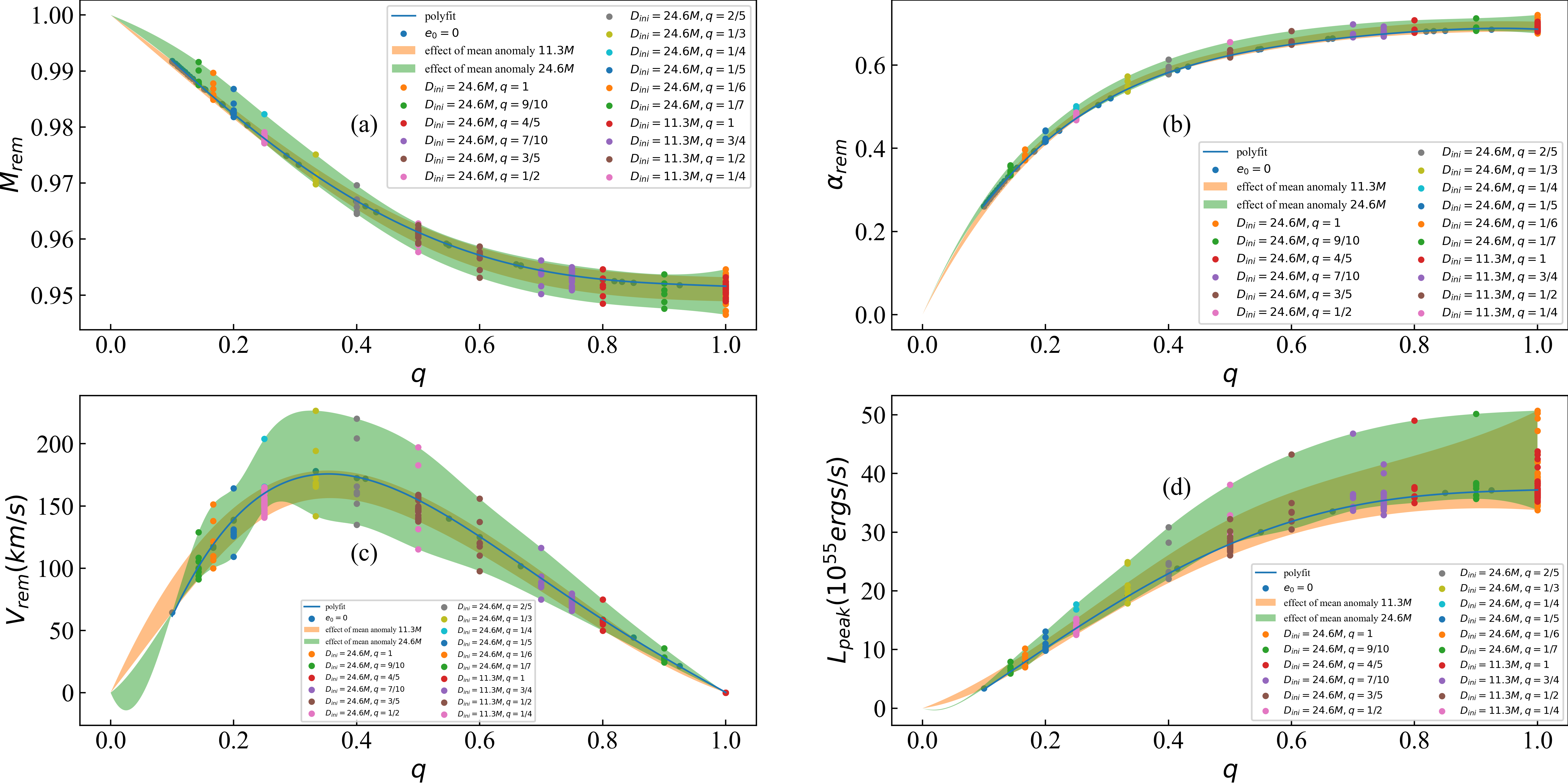}
\caption{\label{FIG:11}Summary of the relationships between the dynamical quantities $M_{\text{rem}}$ (a), $\alpha_{\text{rem}}$ (b), $V_{\text{rem}}$ (c), and $L_{\text{peak}}$ (d) and the mass ratio, combining eccentric and circular orbits for orbital BBH mergers. We interpolate the maximum and minimum values (i.e., boundary value) of the dynamical quantities concerning the mass ratio for two distinct initial distances ($11.3M$ and $24.6M$) of orbital BBH mergers to demonstrate the effect of mean anomaly and incorporate scattered points representing circular orbit dynamics and their polynomial fits for a more intuitive comparison. The scattered points in the figure, labeled by different initial separations and mass ratios, correspond to individual eccentric simulations. `$e_0=0$' denotes circular-orbit NR simulations. `polyfit’ indicates the polynomial fit to the circular-orbit dynamical quantities, while the orange and green shaded regions, defined by the boundaries of the $11.3M$ and $24.6M$ NR simulations, respectively, illustrate the influence of the mean anomaly.}
\end {figure*}

\begin{figure*}[htbp!]
\centering
\includegraphics[width=16cm,height=8cm]{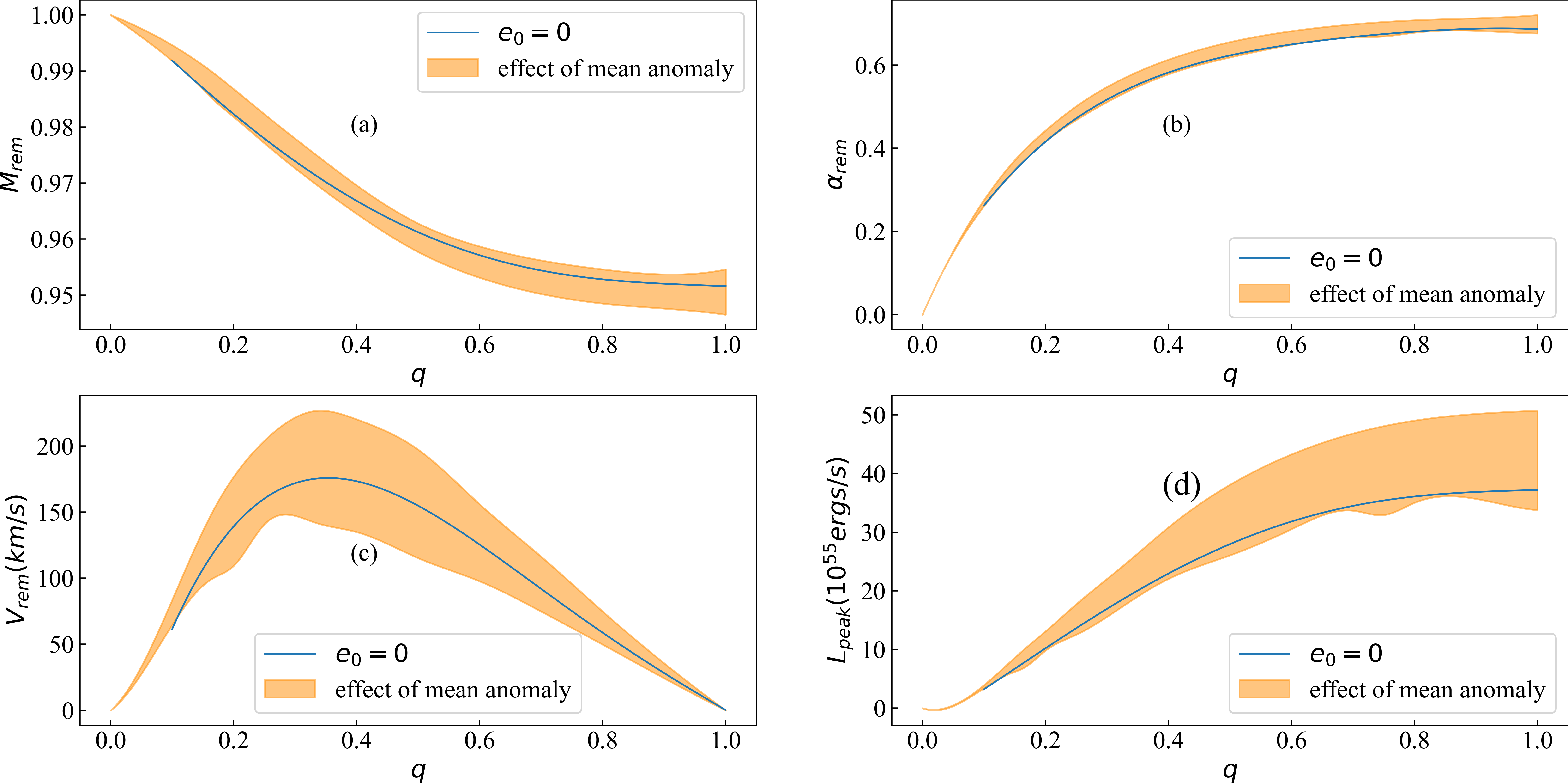}
\caption{\label{FIG:12}Vicinities of dynamical quantities $M_{\text{rem}}$ (a), $\alpha_{\text{rem}}$ (b), $V_{\text{rem}}$ (c), and $L_{\text{peak}}$ (d) derived from the combination of data points for $11.3M$ and $24.6M$, which are formed by the effect of the mean anomaly. `$e_0=0$' represent the polynomial fit of the circular orbit dynamical quantities, and the orange shaded areas are formed by the boundaries combining the dynamical quantities of $11.3M$ and $24.6M$.}
\end {figure*}

\section{Effect of mean anomaly}\label{sec:III}
\subsection{Envelope formation}\label{sec:III:A}
The initial mean anomaly $l_0$ in both NR and PN waveforms discussed in FIG. \ref{FIG:6} remains constant. While we are unable to modify $l_0$ in NR waveforms, we can vary $l_0$ in PN waveforms fitted to NR waveforms during the inspiral phase for FIG. \ref{FIG:6}. This adjustment enables us to compute dynamical quantities such as radiated energy $E_{\mathrm{rad}}$, radiated angular momentum $L_{\mathrm{rad}}$, and radiated linear momentum $P_{\mathrm{rad}}$ using Eqs. (\ref{eq:3}), (\ref{eq:5}), and (\ref{eq:6}) for all possible values of $l_0$.
The eccentric orbital waveforms from RIT correspond to specific initial $l_0$ values. However, by manipulating $l_0$, the resultant calculated radiative quantities correspond to distinct values, preserving the original ${e_t}_0$ and $x_0$ parameters but with a different $l_0$. As we systematically vary $l_0$ across the range from 0 to $2\pi$, we explore all feasible values of the radiative quantities, capturing a specific range of outcomes.

FIG. \ref{FIG:9} illustrates the values and ranges of radiative quantities—radiated energy $E_{\mathrm{rad}}$, radiated angular momentum $L_{\mathrm{rad}}$, and radiated linear momentum $P_{\mathrm{rad}}$, which are calculated from PN waveforms fitted to NR waveforms during the inspiral phase ($200M$ before the merger) for specific mass ratios, as the initial mean anomaly $l_0$ varies continuously from 0 to $2\pi$. The initial eccentricity $e_0$ from RIT is utilized here to maintain consistency with previous works \cite{Wang:2024afj}. 

In panels (a) and (b) of FIG. \ref{FIG:9}, we present the relationship between radiated energy $E_{\mathrm{rad}}$ and radiated angular momentum $L_{\mathrm{rad}}$ calculated using PN waveforms with a mass ratio of $q=1$, and the initial eccentricity $e_0$ in FIG. \ref{FIG:6}. The green line representing the PN (2,2) mode illustrates the oscillatory trend originally shown as scatter points in FIG. \ref{FIG:6}. The light blue scatter points depict the results obtained by varying the $l_0$ value continuously around each original PN fitting scatter point within the range [0, $2\pi$] for 20 consecutive points. The maximum and minimum lines denote the extreme values among these scatter points. 

In panel (c) of FIG. \ref{FIG:9}, we depict the relationship between radiated linear momentum $P_{\mathrm{rad}}$, computed using PN waveforms with mass ratio of $q=1/2$, and the initial eccentricity $e_0$ in FIG. \ref{FIG:6}. The representation of lines and points in panel (c) mirrors that of panels (a) and (b). We specifically employ a mass ratio of $q=1/2$ here, as the radiated linear momentum is zero when $q=1$ due to symmetry considerations. Furthermore, our calculation results also incorporate $\ell \leq 4$ in Eq. (\ref{eq:5}).

In FIG. \ref{FIG:9}, each vertical points corresponds to a specific initial eccentricity $e_0$. However, as $l_0$ continuously changes, these points progressively populate the space between the maximum and minimum bounds. This observation highlights that the waveforms fitted in FIG. \ref{FIG:6} represent merely a subset of $l_0$ values within the defined range, and the oscillation of radiated energy and angular momentum with initial eccentricity is a particular manifestation of this interplay. Notably, the continuous variation of $l_0$ is primarily aimed at maximizing the exploration of the parameter space $[0,2\pi]$ and does not carry a specific significance.

Contrary to a scenario where adjusting $l_0$ might lead to a linear progression of $E_{\mathrm{rad}}$, $L_{\mathrm{rad}}$, and $P_{\mathrm{rad}}$ from a minimum to maximum at a given eccentricity, our findings demonstrate that continuously changing $l_0$ leads to oscillations in these dynamical quantities for a single eccentricity. While the initial eccentricities displayed in FIG. \ref{FIG:9} may not be densely distributed, resulting in visible gaps between different initial eccentricities, it is foreseeable that denser eccentricity distributions would obscure these gaps, leaving behind regions delineated by the maximum and minimum values.
These regions effectively serve as envelopes encapsulating all variations stemming from alterations in $l_0$ and $e_0$, hence we call it  ``envelope of the radiative quantities.'' The envelopes depicted in FIG. \ref{FIG:9} exhibit similarities and distinctions. Notably, as eccentricity rises, the envelope's span widens and accompanied by oscillations, signifying a heightened effect of eccentricity and mean anomaly on these quantities. Both radiative energy and angular momentum envelopes display a consistent increase, while the radiative linear momentum envelope showcases irregularities with a larger range compared to the energy and angular momentum envelopes. Furthermore, the energy envelope surpasses that of the angular momentum.

The emergence of ``Oscillations'' in the radiative quantities displayed in FIG. \ref{FIG:9} hinges on the presence of NR simulations with a fixed initial distance and a sufficiently broad array of continuously varying initial eccentricities. The oscillations in radiative quantities are fundamentally driven by the continuous adjustment of initial eccentricities in NR simulation. In scenarios where the number of NR simulations is limited, such oscillatory patterns may remain imperceptible. 
RIT has conducted a series of BBH merger simulations involving eccentric orbits, with two distinct initial distances: $D_{\text{ini}}=11.3 M$ and $D_{\text{ini}}=24.6 M$. Among these simulations, only a subset featuring continuously changing eccentricities in five cases vividly exhibit the oscillatory behavior of dynamical quantities. In contrast, other cases lack observable oscillations due to the limited number of simulations.
The five cases corresponding to specific initial distances and mass ratios are as follows: $D_{\text{ini}}=24.6 M$ with $q=1$, $D_{\text{ini}}=11.3 M$ with $q=1$, $D_{\text{ini}}=11.3 M$ with $q=3/4$, $D_{\text{ini}}=11.3 M$ with $q=1/2$, and $D_{\text{ini}}=11.3 M$ with $q=1/4$. FIG. \ref{FIG:10} illustrates the envelopes of radiative quantities delineated by these five cases with the initial eccentricity for varying $l_0$ in $[0,2\pi]$, reflecting the characteristics elucidated in FIG. \ref{FIG:9}. Notably, in the case of $D_{\text{ini}}=24.6 M$, the prolonged inspiral phase of the BBHs results in longer waveforms and larger radiative quantities compared to the scenario with $D_{\text{ini}}=11.3 M$. Given that linear momentum radiation is null for mass ratio of $q=1$, it is represented solely as a line in panel (c) of FIG. \ref{FIG:10}.

As the radiative quantities form envelopes, the dynamical quantities $\alpha_{\text{rem}}$, $M_{\text{rem}}$, and $V_{\text{rem}}$ also encapsulate similar envelope-like structures. The observed oscillations highlighted in Refs. \cite{Wang:2023vka,Wang:2024afj} are therefore special cases arising from special initial conditions. If the initial mean anomaly is traversed in $[0,2\pi]$, their dynamical quantities will also form an envelope analogous to that of FIG. \ref{FIG:10}, and the peak luminosity will also form the same envelope, although it represents the moment of merger.

\subsection{Vicinity relative to circular orbits}\label{sec:III:B}
In the Sec. \ref{sec:III:B}, we examined the relationship between the variation of radiative quantities $E_{\mathrm{rad}}$, $L_{\mathrm{rad}}$, and $P_{\mathrm{rad}}$ with initial eccentricity. We noted that for BBH simulation series with the same initial distance and mass ratio, the variation of radiative quantities with eccentricity should form an envelope, with oscillation being a special case of this envelope. Consequently, the dynamical quantities $M_{\text{rem}}$, $\alpha_{\text{rem}}$, $V_{\text{rem}}$, and $L_{\text{peak}}$ will also form envelopes as eccentricity changes, given their intricate connection to the radiative quantities.

In FIG. \ref{FIG:3}, we see that the magnitude of oscillations intensifies with higher initial eccentricities. Once these quantities reach their maximum or minimum values, constrained by the initial distance, the motion of the BBH transitions from an orbital merger (where the orbital cycle exceeds 1) to a non-orbital merger (where the orbital cycle is less than 1). During the non-orbital merger phase, the dynamical quantities progressively converge from peak values to stable values or zero as initial eccentricity varies. The influence of the mean anomaly is an orbital effect associated with periodic motion; therefore, we focus solely on the orbital merger in this context.

In FIG. \ref{FIG:11}, we present a summary of the relationships between the dynamical quantities $M_{\text{rem}}$, $\alpha_{\text{rem}}$, $V_{\text{rem}}$, and $L_{\text{peak}}$ across varying mass ratio, combining data from both eccentric and circular orbits for orbital BBH mergers. 
The data points in FIG. \ref{FIG:11} include information from circular orbits ($e_0=0$) with mass ratios ranging from 1/10 to 1, as well as from eccentric orbits ($e_0\neq0$) with initial distances of $11.3M$ and $24.6M$ and mass ratios from 1/7 to 1, as outlined in the parameter space depicted in FIG. \ref{FIG:1}. Although there are no pronounced oscillations in the dynamical quantities for cases other than $q=1$ at an initial distance of $24.6M$, the sparse oscillations that do exist still delineate the range of changes in the dynamical quantities. 

From FIG. \ref{FIG:11}, we see that for orbital BBH mergers, both eccentricity and mean anomaly induce dynamical quantities to span a range, as indicated by the vertical scatter points relative to circular orbits. For a fixed mass ratio, these quantities oscillate around the circular-orbit values, with larger initial eccentricities producing more pronounced deviations. 
As the mean anomaly and initial eccentricity change continuously, these scatter points transition from discrete points to continuous variations, forming a vicinity relative to the circular orbital dynamical quantities for a given mass ratio. This effect intensifies with larger mass ratios for $M_{\text{rem}}$, $\alpha_{\text{rem}}$, and $L_{\text{peak}}$, reaching a maximum value for a mass ratio of approximately 1/3 in the case of $V_{\text{rem}}$, thereby shaping the impact of mass ratio on eccentric orbital BBH mergers.

Nevertheless, given the current limitations imposed by initial distances and the number of NR simulations, the vicinities delineated by mean anomaly and eccentricity in FIG. \ref{FIG:11} represent only the maximum extent of the effects observed in NR studies to date. The actual vicinities are expected to expand with increasing initial distance. This expansion is attributed to the ability of larger initial distances to accommodate higher initial eccentricities during the transition from orbital to non-orbital mergers.

For each mass ratio presented in FIG. \ref{FIG:11}, the dynamical quantities associated with eccentric orbits form a vicinity relative to those of circular orbits, and it is expected that other mass ratios will exhibit similar behavior. However, there is currently no corresponding NR simulation data, and conducting a large number of simulations would inevitably require substantial computational resources. Nonetheless, the simulation series depicted in FIG. \ref{FIG:11} reveals discernible patterns.
We can interpolate the maximum and minimum values (i.e., boundary values) of the dynamical quantities in relation to the mass ratio for two distinct initial distances of orbital BBH mergers: $11.3M$ and $24.6M$. During this interpolation process, we incorporate points where the dynamical quantities are established at the limit of $q=0$, specifically reflected as $M_{\text{rem}}=1$, $\alpha_{\text{rem}}=0$, $V_{\text{rem}}=0$, and $L_{\text{peak}}=0$. Notably, at these limit points, the vicinity created by eccentricity and mean anomaly vanishes, resulting in the coincidence of the maximum and minimum values.
Occasionally, the dynamical quantity for the circular orbit may exceed the interpolated vicinity at $q=1/10$, situated on the boundary. In such cases, it is essential to include the value of $q=1/10$ as either the maximum or minimum in the interpolation to maintain physical consistency, as the dynamical quantity for the circular orbit must be encompassed within the vicinity.

By comparing the interpolated vicinities for the two distinct initial distances, we observe that the vicinity shaped by $24.6M$ encompasses the vicinity delineated by $11.3M$. This indicates that a larger initial distance results in a broader boundary, consistent with our earlier analysis that higher initial eccentricity induces stronger oscillations.
Additionally, in FIG. \ref{FIG:11}, we incorporate scattered points representing the dynamics of circular orbits along with their polynomial fits for a more intuitive comparison. Notably, the vicinities significantly exceed the extent captured by the polynomial fit.

As previously noted, the data points for $24.6M$ are not densely distributed, which can result in instances where the data points for $11.3M$ fall outside the range of $24.6M$. To establish an upper limit for the vicinity influenced by the effects of eccentricity and mean anomaly, it is essential to synthesize the outcomes from both cases.

In FIG. \ref{FIG:12}, we present the vicinities derived from the combination of data points for $11.3M$ and $24.6M$, shaped by the effects of the mean anomaly. This combination involves determining the maximum or minimum values for the same mass ratio from both initial distances, selecting the larger or smaller value as appropriate.
Additionally, when encountering data points that are notably sparse and exist solely for $11.3M$ but not for $24.6M$ (e.g., at $q=3/4$ and $D_{\text{ini}}=11.3M$), these points are excluded. This approach ensures the derivation of a more rational and comprehensive vicinity, thereby enhancing the overall integrity of the analysis.

In FIG. \ref{FIG:12}, the vicinities in dynamical quantities relative to the circular orbit induced by the effect of eccentricity and mean anomaly are displayed, which is the upper limit of the current NR simulation. This is due to the finite initial distance of the simulation, which leads to the finite initial eccentricity. In the future, there will be more comprehensive simulations of eccentric BBH mergers to improve this upper limit. 

\begin{figure*}[htbp!]
\centering
\includegraphics[width=16cm,height=8cm]{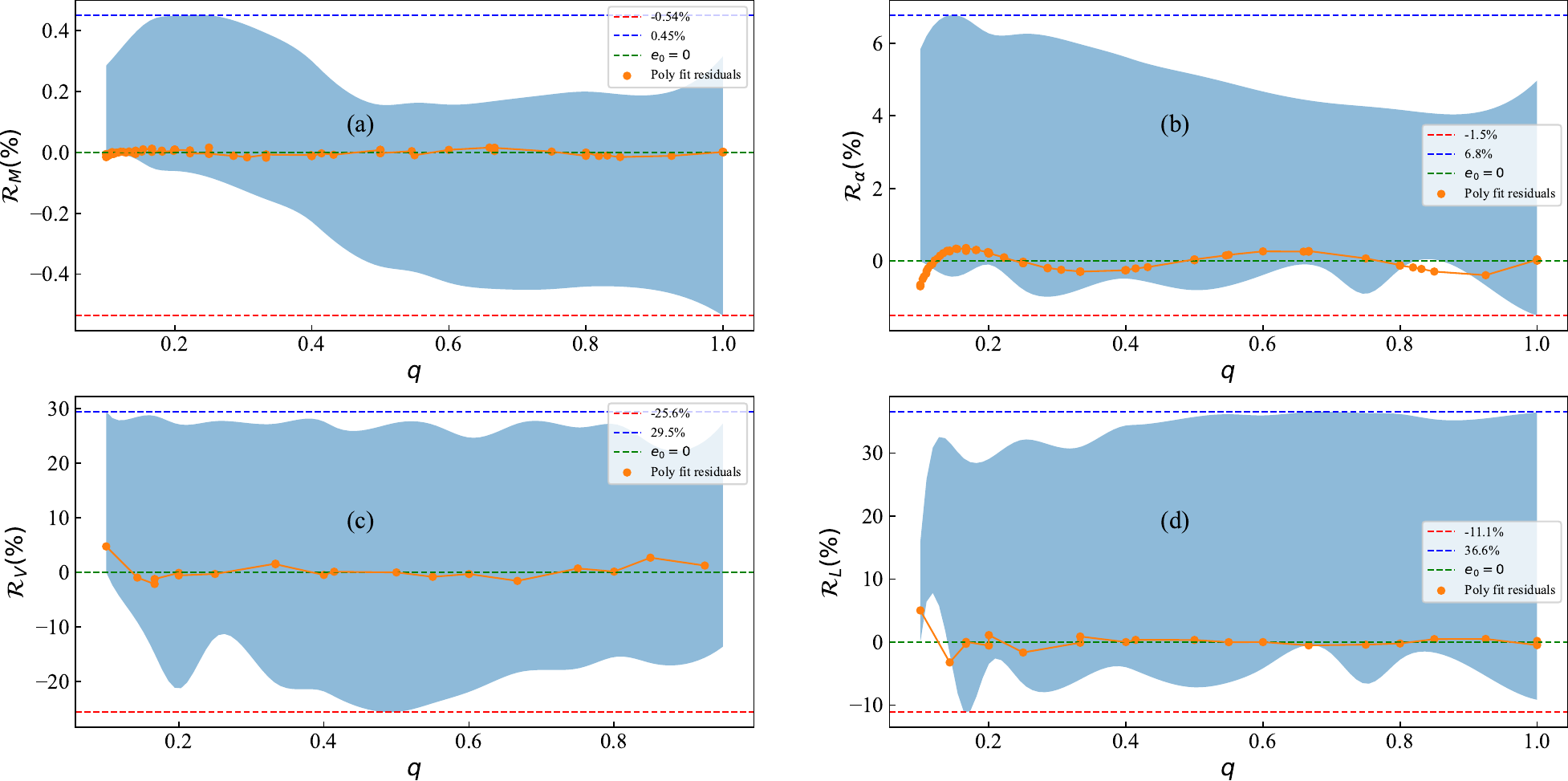}
\caption{\label{FIG:13}Percentage residual of the dynamical quantities $M_{\text{rem}}$ (a), $\alpha_{\text{rem}}$ (b), $V_{\text{rem}}$ (c), and $L_{\text{peak}}$ (d) resulting from the effect of mean anomaly relative to the dynamical quantities of the circular orbit. The blue shaded region and orange scatter points denote, respectively, the percentage residuals of the vicinities and the polynomial fits of circular orbits relative to the circular-orbit interpolation. The maximum positive and minimum negative residuals are indicated by the blue and red dotted lines.}
\end {figure*}

It is essential to quantitatively characterize the residual deviations of these vicinities compared to the circular orbit and to elucidate how these residuals diverge from those obtained through polynomial modeling of the dynamical quantities of the circular orbit. In FIG. \ref{FIG:13}, we employ the percentage residual defined as:
\begin{equation}\label{eq:20}
\mathcal{R}=\frac{A_{\text{e}}-A_\text{c}}{A_\text{c}} \times 100 \% ,  
\end{equation} 
to compute the residuals of the dynamical quantities $M_{\text{rem}}$, $\alpha_{\text{rem}}$, $V_{\text{rem}}$, and $L_{\text{peak}}$ resulting from the effect of mean anomaly relative to the dynamical quantities of the circular orbit, where, $A_\text{e}$ represents the interpolation of the dynamical quantities for eccentric orbits, while $A_\text{c}$ denotes the interpolation for circular orbits. This computation is based on the boundaries of these vicinities, specifically the maximum and minimum values. For comparison, we also include the residuals derived from polynomial modeling of circular orbits.

In FIG. \ref{FIG:13}, it is evident that the residuals arising from the effects of mean anomaly and eccentricity on the BBH merger are significantly larger than those derived from polynomial modeling of the circular orbit. For instance, in panel (a), the largest positive residuals for $M_{\text{rem}}$ reach up to 0.45\%, while the smallest negative residuals extend to -0.54\%. These positive and negative residuals occur because the dynamical quantities in these vicinities exceed or fall below the corresponding circular orbit data.
In panels (b), (c), and (d) of FIG. \ref{FIG:13}, the largest positive residuals for $\alpha_{\text{rem}}$, $V_{\text{rem}}$, and $L_{\text{peak}}$  peak at 6.8\%, 29.5\%, and 36.6\%, respectively, with the smallest negative residuals reaching -1.5\%, -25.6\%, and -11.1\%. The residuals introduced by these vicinities are approximately 25 times, 10 times, 6 times, and 7 times larger than those from polynomial modeling for $M_{\text{rem}}$, $\alpha_{\text{rem}}$, $V_{\text{rem}}$, and $L_{\text{peak}}$, respectively. This stark contrast is highlighted by the substantial differences between the maximum and minimum values of the residuals compared to those of polynomial modeling in FIG. \ref{FIG:13}. This also means that this effect cannot be ignored in the process of dynamic quantity modeling.

\section{Conclusion and Outlook}\label{sec:IV}
In a previous study, we identified universal oscillations in dynamical quantities peak luminosity $L_{\text{peak}}$, remnant mass $M_{\text{rem}}$, remnant spin $\alpha_{\text{rem}}$, and recoil velocity $V_{\text{rem}}$ as functions of initial eccentricity. We correlated these oscillations with integer orbital cycles numbers in a phenomenological framework. In this work, we aim to uncover the physical nature of the oscillations from the perspective of gravitational waveforms. We demonstrate that these oscillations arise from the effects of the mean anomaly and explore how it impacts the dynamical quantities of eccentric BBH mergers.

Focusing on remnant mass and spin, we demonstrate that, while the initial ADM mass ($M_{\mathrm{ADM}}$) and initial orbital angular momentum ($L_0$) vary smoothly with initial eccentricity, the radiated energy ($E_{\text{rad}}$) and angular momentum ($L_{\text{rad}}$) exhibit oscillations that directly correspond to those observed in $M_{\text{rem}}$ and $\alpha_{\text{rem}}$.

Employing 3PN methods to replicate the waveform and its radiative properties during the inspiral phase and comparing the radiated energy and angular momentum from NR waveforms, PN waveforms, and orbital-averaged PN fluxes, we find that the mean anomaly $l_0$ as the source of the oscillations. The effect of $l_0$ is averaged out in orbital-averaged flux calculations. Utilizing eccentric PN waveforms to calculate dynamical quantities, we find that this effect significantly impacts the mass, spin, and recoil velocity of the merger remnant, with its influence increasing as the initial eccentricity rises. Specifically, in the PN framework, the amplitudes of oscillations for mass, spin, and recoil velocity at ${e_t}_0 = 0.5$ are enhanced by approximately 10, 5, and 7 times, respectively, compared to those at ${e_t}_0 = 0.1$. For a circular orbit, where ${e_t}_0 = 0.0$, the oscillations vanish entirely.

While $l_0$ in NR waveforms remains fixed, we explore the flexibility of varying $l_0$ in PN waveforms fitted to NR waveforms during the inspiral phase. This adjustment enables us to calculate radiative quantities such as radiated energy, angular momentum, and linear momentum for various $l_0$ values. We find that by continuously varying the mean anomaly $l_0$ within the parameter space $[0,2\pi]$ from a PN perspective, we can create an envelope that encapsulates the original oscillations of these radiative quantities. This observation indicates that the oscillations arise from the specific initial condition $l_0$, while the actual mean anomaly influences the formation of the envelope.

Analyzing the relationships between dynamical quantities and mass ratio for eccentric and circular orbital BBH mergers, our findings underscore the interplay of eccentricity and mean anomaly in inducing oscillations and ranges in dynamical quantities relative to circular orbits, influenced by the continuous changes in initial eccentricity and the associated envelope. In the context of orbital mergers, we interpolate the maximum and minimum values of the dynamical quantities to delineate the vicinities of dynamical quantities for eccentric BBH mergers. We find that the impact of mean anomaly intensifies with higher mass ratios for $M_{\text{rem}}$, $\alpha_{\text{rem}}$, and $L_{\text{peak}}$, and reach maximum effects for a mass ratio of approximately 1/3 on $V_{\text{rem}}$. 

By quantifying residual deviations compared to circular orbit dynamics, we highlight substantial differences between the effects of mean anomaly and eccentricity on BBH mergers and polynomial modeling of circular orbits. Residuals stemming from these effects are notably larger for dynamical quantities of eccentric orbits than circular orbits, emphasizing it is an important effect that cannot be ignored.
 
The vicinities of dynamical quantities shaped by mean anomaly in relation to circular orbits are derived from all current eccentric BBH simulations conducted by RIT, and they are not exhaustive. These vicinities represent the upper limits of our current understanding, and we anticipate that their extents will expand with increased initial distances in eccentric NR simulations. In this study, for waveforms with $e_0=0.2$ at different initial separations, the $11.3M$ waveform lasts $1000M$, whereas the $24.6M$ waveform extends to nearly $12000M$. Doubling the initial separation increases the evolution time by roughly an order of magnitude, with even longer durations for smaller mass ratios. The initial eccentricity at the oscillation peak reaches 0.6 for the $24.6M$ case, compared with 0.3 for 
$11.3M$. Therefore, achieving stronger oscillations requires both higher eccentricities and larger initial separations, which can increase the computational cost by nearly an order of magnitude relative to the 
$24.6M$ case. In the future, with sufficient computational resources, simulations at larger separations and smaller mass ratios could be performed, enabling coverage of vicinities with mean anomalies across the full eccentricity and mass ratio range.
As these vicinities develop, our comprehension of eccentric BBH mergers will improve, with future simulations at larger initial distances likely to refine the boundaries of these vicinities. Moreover, further simulations incorporating spin alignment and spin precession configurations will enhance our understanding of eccentric orbital BBH mergers.

The findings outlined in our study offer a detailed exploration of the interplay between eccentricity, mean anomaly, and dynamical quantities in BBH mergers. Through thorough numerical simulations and analyses, we uncover the effect of the mean anomaly on eccentric orbital BBH mergers, providing valuable insights for further investigations in gravitational wave astronomy.

\textbf{Astrophysical Implications:} The effects of eccentricity and mean anomaly primarily impact the amplitude of gravitational waves, particularly the peak amplitude, which closely corresponds to the peak luminosity. As shown in FIG. \ref{FIG:13}, these effects can induce a maximum relative deviation of 36.6\%. Neglecting them, even when using current eccentric orbital BBH waveforms, can lead to biases in the signal-to-noise ratio during matched filtering, affecting assessments of signal significance. In parameter estimation, amplitude deviations can systematically under or overestimate luminosity distances, alter inferred source inclination and polarization angles, and influence merger rate inferences. Therefore, incorporating the influence of the mean anomaly into gravitational-wave models is crucial for improving data analysis. From an astrophysical perspective, neglecting this effect can introduce significant errors in estimates of remnant black-hole mass, spin, and recoil, impacting our understanding of black-hole formation channels. Accounting for these effects is thus essential for accurate modeling of BBH mergers and their astrophysical implications.

\begin{acknowledgments}
The authors are very grateful to the RIT and SXS collaboration for the numerical simulation of BBH mergers, and thanks to Yan Fang Huang, Yu Liu and Xiaolin Liu for their helpful discussions. BL acknowledges support from the National Key Research and Development Program of China (No. 2023YFB3002502) and National Natural Science Foundation of China (Grant No. 12433008). This work is supported by the National Key R\&D Program of China (2021YFA0718504).
\end{acknowledgments}

\bibliographystyle{apsrev4-2}

\bibliography{ref}
\appendix

\section{Orbital average energy and angular momentum fluxs}\label{App:A}
\begin{widetext}
According to the Ref. \cite{Arun:2009mc}, in ADM coordinates (we can transform it to harmonic coordinates using Eq. (A45) from Ref.\cite{Wang:2024jro}), the orbital average energy flux of 3PN can be expressed as the sum of the instantaneous term and the hereditary term:
\begin{equation}
\left\langle\mathcal{F}\right\rangle= \left\langle\mathcal{F}_{\text {inst }}\right\rangle + \left\langle\mathcal{F}_{\text {hered }}\right\rangle,
\end{equation}
where $\left\langle\mathcal{F}\right\rangle$ represents the orbital average or periodic average of $\mathcal{F}$, and the instantaneous term and hereditary term are expressed as:
\begin{equation}
\begin{aligned}
&\left\langle\mathcal{F}_{\text {inst }}\right\rangle=\frac{32}{5} \frac{c^5}{G} \eta^2 x^5\left(\mathcal{F}_{\mathrm{Newt}}+\mathcal{F}_{1 \mathrm{PN}} x+\mathcal{F}_{2 \mathrm{PN}} x^2+\mathcal{F}_{3 \mathrm{PN}} x^3\right),\\
&\left\langle\mathcal{F}_{\text {hered }}\right\rangle=\frac{32}{5} \frac{c^5}{G} \eta^2 x^5\left(\mathcal{F}_{1.5 \mathrm{PN}} x^{3 / 2}+\mathcal{F}_{2.5 \mathrm{PN}} x^{5 / 2}+\mathcal{F}_{3 \mathrm{PN}} x^3\right).
\end{aligned}
\end{equation}
The expansion coefficients of the instantaneous part can be expressed as
\begin{equation}
\mathcal{F}_{\mathrm{Newt}}=\frac{1}{\left(1-e_t^2\right)^{7 / 2}}\left\{1+\frac{73}{24} e_t^2+\frac{37}{96} e_t^4\right\},
\end{equation}
\begin{equation}
\begin{aligned}
\mathcal{F}_{1 \mathrm{PN}}=&\frac{1}{\left(1-e_t^2\right)^{9 / 2}}\left\{-\frac{1247}{336}-\frac{35}{12} \eta+e_t^2\left(\frac{10475}{672}-\frac{1081}{36} \eta\right)\right.\\&\left.+e_t^4\left(\frac{10043}{384}-\frac{311}{12} \eta\right)+e_t^6\left(\frac{2179}{1792}-\frac{851}{576} \eta\right)\right\},
\end{aligned}
\end{equation}
\begin{equation}
\begin{aligned}
\mathcal{F}_{2 \mathrm{PN}}= & \frac{1}{\left(1-e_t^2\right)^{11 / 2}}\left\{-\frac{203471}{9072}+\frac{12799}{504} \eta+\frac{65}{18} \eta^2+e_t^2\left(-\frac{3866543}{18144}+\frac{4691}{2016} \eta+\frac{5935}{54} \eta^2\right)\right.\\&\left.+e_t^4\left(-\frac{369751}{24192}-\frac{3039083}{8064} \eta\right.\right.  \left.+\frac{247805}{864} \eta^2\right)+e_t^6\left(\frac{1302443}{16128}-\frac{215077}{1344} \eta+\frac{185305}{1728} \eta^2\right) \\&+e_t^8\left(\frac{86567}{64512}-\frac{9769}{4608} \eta+\frac{21275}{6912} \eta^2\right) \left.+\sqrt{1-e_t^2}\left[\frac{35}{2}-7 \eta+e_t^2\left(\frac{6425}{48}-\frac{1285}{24} \eta\right)\right.\right.\\&\left.\left.+e_t^4\left(\frac{5065}{64}-\frac{1013}{32} \eta\right)+e_t^6\left(\frac{185}{96}-\frac{37}{48} \eta\right)\right]\right\},
\end{aligned}
\end{equation}
\begin{equation}
\begin{aligned}
\mathcal{F}_{3 \mathrm{PN}}= & \frac{1}{\left(1-e_t^2\right)^{13 / 2}}\left\{\frac{2193295679}{9979200}+\left[\frac{8009293}{54432}-\frac{41}{64} \pi^2\right] \eta-\frac{209063}{3024} \eta^2-\frac{775}{324} \eta^3\right.\\&\left.+e_t^2\left(\frac{2912411147}{2851200}+\left[\frac{249108317}{108864}\right.\right.\right.  \left.\left.+\frac{31255}{1536} \pi^2\right] \eta-\frac{3525469}{6048} \eta^2-\frac{53696}{243} \eta^3\right)\\&+e_t^4\left(-\frac{4520777971}{13305600}+\left[\frac{473750339}{108864}-\frac{7459}{1024} \pi^2\right] \eta+\frac{697997}{576} \eta^2\right. \left.-\frac{10816087}{7776} \eta^3\right)\\&+e_t^6\left(\frac{3630046753}{26611200}+\left[-\frac{8775247}{145152}-\frac{78285}{4096} \pi^2\right] \eta+\frac{31147213}{12096} \eta^2-\frac{983251}{648} \eta^3\right) \\
& +e_t^8\left(\frac{21293656301}{141926400}+\left[-\frac{36646949}{129024}-\frac{4059 \pi^2}{4096}\right] \eta+\frac{85830865}{193536} \eta^2-\frac{4586539}{15552} \eta^3\right)\\&+e_t^{10}\left(-\frac{8977637}{11354112}\right.  \left.+\frac{9287}{48384} \eta+\frac{8977}{55296} \eta^2-\frac{567617}{124416} \eta^3\right)+\sqrt{1-e_t^2}\left[-\frac{14047483}{151200}\right.\\&\left.+\left[-\frac{165761}{1008}+\frac{287}{192} \pi^2\right] \eta+\frac{455}{12} \eta^2\right. +e_t^2\left(\frac{36863231}{100800}+\left[-\frac{14935421}{6048}+\frac{52685}{4608} \pi^2\right] \eta\right.\\&\left.+\frac{43559}{72} \eta^2\right) +e_t^4 \left(\frac{759524951}{403200}+\left[-\frac{31082483}{8064}+\frac{41533}{6144} \pi^2\right] \eta\right. \left.\left.+\frac{303985}{288} \eta^2\right)\right.\\&\left.+e_t^6\left(\frac{1399661203}{2419200}+\left[-\frac{40922933}{48384}+\frac{1517}{9216} \pi^2\right] \eta+\frac{73357}{288} \eta^2\right)+e_t^8\left(\frac{185}{48}-\frac{1073}{288} \eta+\frac{407}{288} \eta^2\right)\right] \\
& \left.+\left(\frac{1712}{105}+\frac{14552}{63} e_t^2+\frac{553297}{1260} e_t^4+\frac{187357}{1260} e_t^6+\frac{10593}{2240} e_t^8\right) \ln \left[\frac{x}{x_0} \frac{1+\sqrt{1-e_t^2}}{2\left(1-e_t^2\right)}\right]\right\}.
\end{aligned}
\end{equation}
The expansion coefficients of the hereditary part can be expressed as
\begin{equation}
\mathcal{F}_{1.5 \mathrm{PN}}=4 \pi x^{3 / 2} \varphi\left(e_t\right),
\end{equation}
\begin{equation}
\mathcal{F}_{2.5 \mathrm{PN}}=\pi \left[-\frac{8191}{672} \psi\left(e_t\right)-\frac{583}{24} \eta \zeta\left(e_t\right)\right] \varphi\left(e_t\right),
\end{equation}
\begin{equation}
\mathcal{F}_{3 \mathrm{PN}}=\left[-\frac{116761}{3675} \kappa\left(e_t\right)+\left[\frac{16}{3} \pi^2-\frac{1712}{105} C\right.\right. \left.\left.-\frac{1712}{105} \ln \left(\frac{4 x^{3/2}}{x_0}\right)\right] F\left(e_t\right)\right],
\end{equation}
where $C = 0.577$, denotes the Euler constant, and the logarithmic term of $x_0$ exists in both the instantaneous term and the hereditary term and cancels each other out, so $x_0$ can be set to an arbitrary constant. $\varphi\left(e_t\right)$, $\psi \left(e_t\right)$, $\zeta \left(e_t\right)$, $\kappa \left(e_t\right)$ and $F \left(e_t\right)$ represent some special functions, among which only $F (e_t)$ has an analytical form, which can be expressed as
\begin{equation}
F(e_t)=\frac{1+\frac{85}{6} {e_t}^2+\frac{5171}{192} {e_t}^4+\frac{1751}{192} {e_t}^6+\frac{297}{1024} {e_t}^8}{\left(1-{e_t}^2\right)^{13 / 2}}.
\end{equation}
The exact results for the other four special functions can be obtained by interpolating the data in Appendix B of Ref. \cite{Arun:2009mc}.

Similarly, in ADM coordinates, the orbital average  angular momentum flux of 3PN can be expressed as the sum of the instantaneous term and the hereditary term:
\begin{equation}
\left\langle\mathcal{G}\right\rangle= \left\langle\mathcal{G}_{\text {inst }}\right\rangle + \left\langle\mathcal{G}_{\text {hered }}\right\rangle,
\end{equation}
where $\left\langle\mathcal{G}\right\rangle$ represents the orbital average of $\mathcal{G}$, and the instantaneous term and hereditary term are expressed as:
\begin{equation}
\begin{aligned}
\left\langle\mathcal{G}_{\text {inst }}\right\rangle&=\frac{4}{5} c^2 m \eta^2 x^{7 / 2}\left(\mathcal{G}_{\mathrm{Newt}}+x \mathcal{G}_{1 \mathrm{PN}}+x^2 \mathcal{G}_{2 \mathrm{PN}}+x^3 \mathcal{G}_{3 \mathrm{PN}}\right),\\
\left\langle\mathcal{G}_{\text {hered }}\right\rangle&=\frac{32}{5} {c^2}m \eta^2 x^{7/2}\left(\mathcal{G}_{1.5 \mathrm{PN}} x^{3 / 2}+\mathcal{G}_{2.5 \mathrm{PN}} x^{5 / 2}+\mathcal{G}_{3 \mathrm{PN}} x^3\right).
\end{aligned}
\end{equation}
The expansion coefficients of the instantaneous part can be expressed as
\begin{equation}
\mathcal{G}_{\mathrm{Newt}}=\frac{8+7 e_t^2}{\left(1-e_t^2\right)^2},
\end{equation}
\begin{equation}
\mathcal{G}_{1 \mathrm{PN}}=\frac{1}{\left(1-e_t^2\right)^3}\left\{-\frac{1247}{42}-\frac{70}{3} \eta+\left(\frac{3019}{42}-\frac{335}{3} \eta\right) e_t^2+\left(\frac{8399}{336}-\frac{275}{12} \eta \right) e_t^4\right\},
\end{equation}
\begin{equation}
\begin{aligned}
\mathcal{G}_{2 \mathrm{PN}}&=\frac{1}{\left(1-e_t^2\right)^4}\left\{-\frac{135431}{1134}+\frac{11287}{63} \eta+\frac{260}{9} \eta^2+\left(-\frac{607129}{756}-\frac{6925}{84} \eta+\frac{1546}{3} \eta^2\right) e_t^2\right.\\
& +\left(\frac{28759}{432}-\frac{116377}{168} \eta+569 \eta^2\right) e_t^4+\left(\frac{30505}{2016}-\frac{2201}{56} \eta+\frac{1519}{36} \eta^2\right) e_t^6 \\
& \left.+\sqrt{1-e_t^2}\left[80-32 \eta+(335-134 \eta) e_t^2+(35-14 \eta) e_t^4\right]\right\},
\end{aligned}
\end{equation}
\begin{equation}
\begin{aligned}
\mathcal{G}_{3 \mathrm{PN}}= & \frac{1}{\left(1-e_t^2\right)^5}\left\{\frac{2017023341}{1247400}+\frac{4340155}{6804} \eta-\frac{167483}{378} \eta^2-\frac{1550}{81} \eta^3\right. \\
& +e_t^2\left(\frac{153766369}{44550}+\left[\frac{15157061}{1701}+\frac{647}{8} \pi^2\right] \eta-\frac{116335}{54} \eta^2-\frac{96973}{81} \eta^3\right) \\
& +e_t^4\left(-\frac{6561101941}{1663200}+\left[\frac{163935875}{18144}-\frac{6817}{256} \pi^2\right] \eta+\frac{3541255}{1008} \eta^2-\frac{438907}{108} \eta^3\right) \\
& +e_t^6\left(-\frac{10123087}{19800}+\left[-\frac{326603}{2268}-\frac{615}{128} \pi^2\right] \eta+\frac{2224003}{1008} \eta^2-\frac{283205}{162} \eta^3\right) \\
& +e_t^8\left(-\frac{10305073}{709632}+\frac{417923}{12096} \eta+\frac{95413}{8064} \eta^2-\frac{146671}{2592} \eta^3\right) \\
& +\sqrt{1-e_t^2}\left[-\frac{379223}{630}+\left[-\frac{48907}{63}+\frac{41}{6} \pi^2\right] \eta+\frac{580}{3} \eta^2\right. \\
& +e_t^2\left(\frac{309083}{315}+\left[-\frac{456250}{63}+\frac{2747}{96} \pi^2\right] \eta+1902 \eta^2\right) \\
& +e_t^4\left(\frac{13147661}{5040}+\left[-\frac{2267795}{504}+\frac{287}{96} \pi^2\right] \eta+\frac{2703}{2} \eta^2\right)  \left.+e_t^6\left(70-\frac{203}{3} \eta+\frac{77}{3} \eta^2\right)\right] \\
& \left.+\left(\frac{13696}{105}+\frac{98012}{105} e_t^2+\frac{23326}{35} e_t^4+\frac{2461}{70} e_t^6\right) \ln \left[\frac{x}{x_0} \frac{1+\sqrt{1-e_t^2}}{2\left(1-e_t^2\right)}\right]\right\}.
\end{aligned}
\end{equation}
The expansion coefficients of the hereditary part can be expressed as
\begin{equation}
\mathcal{G}_{1.5 \mathrm{PN}} = 4 \pi \tilde{\varphi}\left(e_t\right),
\end{equation}
\begin{equation}
\mathcal{G}_{2.5 \mathrm{PN}} = \pi\left[-\frac{8191}{672} \tilde{\psi}\left(e_t\right)-\frac{583}{24} \eta \tilde{\zeta}\left(e_t\right)\right],
\end{equation}
\begin{equation}
\mathcal{G}_{3 \mathrm{PN}} = -\frac{116761}{3675} \tilde{\kappa}\left(e_t\right)+\left[\frac{16}{3} \pi^2-\frac{1712}{105} C-\frac{1712}{105} \ln \left(\frac{4 c x^4}{x_0}\right) \tilde{F}\left(e_t\right)\right],
\end{equation}
where $\tilde{\varphi}\left(e_t\right)$, $\tilde{\psi} \left(e_t\right)$, $\tilde{\zeta} \left(e_t\right)$, $\tilde{\kappa} \left(e_t\right)$ and $\tilde{F} \left(e_t\right)$ represent some special functions, among which only $\tilde{F} (e_t)$ has an analytical form, which can be expressed as
\begin{equation}
\tilde{F}(e_t)=\frac{1+\frac{229}{32} {e_t}^2+\frac{327}{64} {e_t}^4+\frac{69}{256} {e_t}^6}{\left(1-{e_t}^2\right)^5}.
\end{equation}
The exact results for the other four special functions can be obtained by interpolating the data in Appendix B of Ref. \cite{Arun:2009mc}.

\end{widetext}

\end{document}